\documentclass[aps,groupedaddress,superscriptaddress,amsmath,amssymb,twocolumn,pre]{revtex4-2}
\usepackage[dvips]{graphicx}
\usepackage{latexsym}
\usepackage{amsthm}
\usepackage{amsfonts}
\usepackage{bm}

\usepackage[colorlinks=true,citecolor=blue,linkcolor=blue,urlcolor=blue, linktocpage=true,breaklinks=true,pagebackref=false]{hyperref}
\usepackage{textcomp} 		%\textmu etc
\usepackage{hyperref} 		%clickable links
\usepackage{ifthen} 		% provides \ifthenelse test
\usepackage{xifthen} 		% provides \isempty test
\usepackage{color} 			%colored
\usepackage{soul} 			%highlights with }
\usepackage{lineno}
\usepackage[normalem]{ulem}

\usepackage{lipsum}
\usepackage{siunitx}
\usepackage{xcolor} 
\usepackage{float}
\usepackage{enumitem}

\usepackage{soul}

\usepackage{physics}

\def\0{^{(0)}}
\def\1{^{(1)}}
\def\sgn{\mathrm{sgn}}
\usepackage{braket}

\def\n{{\bm \nabla}}

\newcommand{\uproman}[1]{\uppercase\expandafter{\romannumeral#1}}

\DeclareMathOperator{\JacobiCN}{cn}
\DeclareMathOperator{\JacobiDN}{dn}
\DeclareMathOperator{\JacobiSN}{sn}

\usepackage{cancel}

\begin{document}
\setstcolor{red}

\title{Resonant-force induced symmetry breaking in a quantum parametric oscillator}
\author{D.\,K.\,J. Boneß}
\affiliation{Department of Physics, University of Konstanz, 78464 Konstanz, Germany}
\author{W. Belzig}
\affiliation{Department of Physics, University of Konstanz, 78464 Konstanz, Germany}
\author{M.\,I. Dykman}
\affiliation{Department of Physics and Astronomy, Michigan State University, East Lansing, MI 48824, USA}
%\email{dykmanm@msu.edu}
%

\date{\today}
	
\begin{abstract} %(max. 500 words)

A parametrically modulated oscillator has two opposite-phase vibrational states at half the modulation frequency. An extra force at the vibration frequency breaks the symmetry of the states. The effect can be extremely strong due to the interplay between the force and the quantum fluctuations resulting from the coupling of the oscillator to a thermal bath. The force changes the rates of the fluctuation-induced walk over the quantum states of the oscillator. If the number of the states is large, the effect accumulates to an exponentially large factor in the rate of switching between the vibrational states. We find the factor and analyze it in the limiting cases including the prebifurcation regime where the system is close but not too close to the bifurcation point.

\end{abstract}
\maketitle

\section{INTRODUCTION}

Quantum dynamics of parametric oscillators has been attracting increasing interest from both theoretical and experimental perspectives \cite{Kryuchkyan1996,Mirrahimi2014,Goto2016,Bartolo2016,Zhang2017b,Puri2017e,Dykman2018,Goto2019a,Yamamoto2020,Roberts2020,Grimm2020,Roberts2021,Ng2022,Venkatraman2024,Zilberberg2023,Chavez-Carlos2023}. To an extent, this interest comes from new applications of parametric oscillators, in particular in quantum information. In a broader context, such oscillators provide a versatile platform for studying quantum dynamics far from thermal equilibrium and revealing its hitherto unknown aspects, with new features of tunnelling and new collective phenomena being examples. One of the features of the dynamics, which is a part of the motivation of the present paper, is the occurrence  and the signatures of detailed balance in a multistate quantum system.

To a large extent, the importance of parametric oscillators is a consequence of their symmetry. Such oscillators are vibrational systems with periodically modulated parameters (like the eigenfrequency) that display vibrations at half the modulation frequency $\omega_p$. Classically, the vibrational states have equal amplitudes and opposite phases \cite{Landau2004a}, presenting a basic example of period doubling. Quantum mechanically, the vibrational  states can be thought of as generalized coherent states of opposite sign \cite{Mandel1995}. The Floquet eigenstates are symmetric and antisymmetric combinations of vibrational states at frequency $\omega_p/2$.

\begin{figure}[h]
	\centering
	\includegraphics[width=1\linewidth]{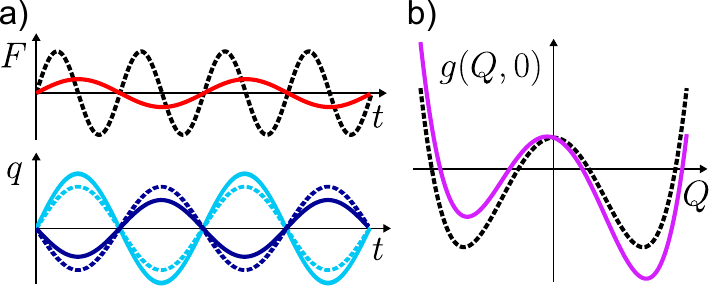}
	\caption{(a) Top panel: sketch of the modulation (dotted line) and the additive extra force (red solid line). Lower panel: Sketch of the vibrating coordinate for the two vibrational states of the oscillator for no extra force (dotted line), the vibrations with the phase close to the phase of the extra force (light blue line), and the vibrations in the counterphase with the force (dark blue line). (b) The cross-section of the scaled Hamiltonian function of the oscillator in the rotating frame   $g(Q,P)$ as a function of the quadrature $Q$ for $P=0$. For no extra force $g(Q,P)$ is symmetric, $g(Q,P)=g(-Q,-P)$ (dotted line). The extra force breaks the symmetry, making one well of $g(Q,P)$ deeper and the other well  shallower (magenta line).}
	\label{fig:sketch}
\end{figure}

Generally, using  parametric oscillators in quantum information requires operations that would break their symmetry, cf.~\cite{Goto2018}. The symmetry breaking can be implemented by applying an extra force at frequency $\omega_p/2$. Classically, the effect of such force can be understood from Fig.~\ref{fig:sketch}~(a). Because the vibrational states have opposite phases, the force can be in phase with one of the two states, increasing its amplitude, while being in counter-phase with the other state and decreasing its amplitude. The states symmetry is thus broken. However, for a weak force this effect is small.

In the present paper we study the effect of a weak extra force at frequency $\omega_p/2$ on a quantum parametric oscillator. The quantum effect is nonperturbative, in some sense, as it changes the nomenclature of the quantum states. Instead of the Floquet states with the eigenvalues defined modulo $\hbar\omega_p$ \cite{Shirley1965,*Zel'dovich1967,*Ritus1967,*Sambe1973}, in the presence of the extra force the eigenvalues are defined modulo $\hbar\omega_p/2$, and even a weak force can strongly change coherent quantum dynamics. As we show, the force can have a strong effect in the presence of dissipation, too. We study this effect where the dynamics involves multiple oscillator states, in which case it is exponentially strong. Also, we consider the case where the oscillator eigenfrequency $\omega_0$ is close to $\omega_p/2$, so that the  parametric modulation at frequency $\omega_p$ that excites the vibrations can be relatively weak.

Besides the  discreteness of the eigenvalue spectrum, a qualitative distinction between the quantum and classical dynamics comes from the nature of the fluctuations associated with the coupling of the oscillator to a thermal bath. Along with classical thermal fluctuations, the coupling leads to quantum fluctuations. In quantum terms, oscillator relaxation comes from transitions between the oscillator states with emission of excitations into the bath. The emission rate determines the relaxation rate, but the very emission events happen at random, leading to noise. In a modulated oscillator, such noise is present even if the bath temperature is $T=0$.

Quantum and classical fluctuations can strongly enhance the effect of the symmetry breaking by a force at frequency $\omega_p/2$. The effect is ultimately determined by the relation between the appropriately scaled force amplitude and the fluctuation intensity. To provide intuition, we draw an analogy with a Brownian particle in a bistable potential, cf. Fig.~\ref{fig:sketch}~(b). Such analogy is seen if one looks at the vibrating oscillator in the frame rotating at $\omega_p/2$. Here the  dynamics is characterized by the scaled vibration quadratures $Q$ and $P$, i.e., the amplitudes of the vibrational components $\cos(\omega_pt/2)$ and $\sin(\omega_pt/2)$.  These variables can be associated with the scaled coordinate and momentum of the oscillator in the rotating frame. 

With no force at $\omega_p/2$, the rotating-frame Hamiltonian is  even in $\{Q,P\}$ by symmetry: both $\cos(\omega_pt/2)$ and $\sin(\omega_pt/2)$ change sign for $t\to t+2\pi/\omega_p$, whereas the modulation, and thus the Hamiltonian, do not change. The Hamiltonian becomes time-independent in the rotating wave approximation (RWA). Its cross-section  by the plane $P=0$ is sketched in Fig.~\ref{fig:sketch}~(b). It has the form of a double-well potential.  The minima correspond to the stable vibrational states, in the presence of weak dissipation  \cite{Wielinga1993,Marthaler2006}. 

A force at frequency $\omega_p/2$ is seen in the rotating frame as a static bias. It breaks the symmetry of the Hamiltonian. The effect is reminiscent of the effect of bias on a Brownian particle in a symmetric double-well potential. With no bias, the potential wells are equally populated. If the bias changes the well depths by $\delta U$, the rates of thermal-noise-induced interwell switching change \cite{Kramers1940}. As a consequence, the stationary ratio of the well populations changes by $\exp(2\delta U/k_BT)$. This factor can be large for small temperature even where $\delta U$ is small compared to the height of the barrier separating the wells. 

Similar to a static bias for a Brownian particle, a force at $\omega_p/2$ can exponentially strongly affect the rates of noise-induced switching between the vibrational states of a classical parametric oscillator \cite{Ryvkine2006a}. As a result, the stationary populations of the states are also strongly changed. The population change was observed  for a parametrically modulated mode of a micromechanical resonator by Mahboob et al. \cite{Mahboob2010}. Micromechanical resonators were also used by Han et al. \cite{Han2024} to demonstrate a strong characteristic change of the switching rates.

\begin{figure}[h]
	\centering
	\includegraphics[width=1\linewidth]{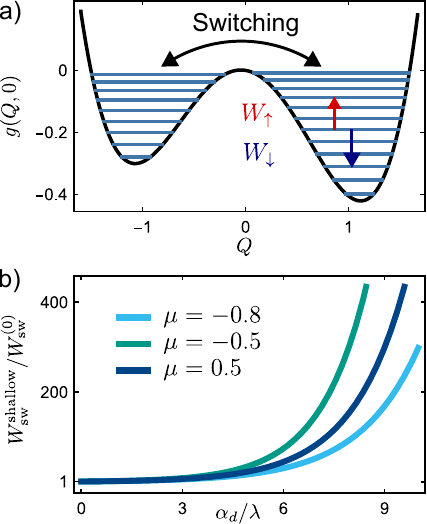}
	\caption{(a) Scaled quasienergy levels of the oscillator for the parameters $\mu = 0.2$, $\lambda = 0.02$, $\alpha_d =0.055$, and $\varphi_d = \pi/2$, see Eqs.~(\ref{eq:full_g}) - (\ref{eq:mu}). $W_\uparrow$ and $W_\downarrow$ indicate transitions due to the coupling to a thermal bath; the transitions are not limited to the nearest quasienergy levels. The rate of transitions down is higher than the rate of transitions up. The states at the bottom of the wells of $g(Q,0)$ correspond to the stable vibrational states.  (b) The strong change of the switching rate $W_\mathrm{sw}$ with the varying scaled amplitude of the extra force $\alpha_d$ compared to the rate $W_\mathrm{sw}^{(0)}$ for $\alpha_d=0$. The plot shows the ratio $W_\mathrm{sw}^\mathrm{shallow}/W_\mathrm{sw}^{(0)}$ for the shallow well. For the deeper well the ratio is inversed.  The data refers to three values of the scaled difference $\mu$ between the oscillator eigenfrequency and half the modulation frequency; $\lambda$ is the scaled Planck constant and the oscillator thermal occupation number is $\bar{n}=0.2$.}
	\label{fig:resonance}
\end{figure}

On the quantum side, of major interest for applications is the regime of comparatively large vibration amplitudes, in which the overlap of the wave functions of the coexisting vibrational states is exponentially small. It corresponds to having many quantum states inside the wells of the scaled RWA Hamiltonian $g(Q,P)$ in Fig.~\ref{fig:resonance}. In this case, similar to the classical regime, oscillator relaxation is characterized by two strongly different rates. One is the decay rate in the absence of modulation which, for a modulated oscillator, determines the time it takes to approach a stable vibrational state at one of the minima of $g(Q,P)$. The other is the rate of switching between the stable vibrational states due to classical and quantum fluctuations, which is exponentially smaller for low fluctuation intensity \cite{Marthaler2006}. 

It is the rate of switching between the vibrational states of a quantum oscillator that can be strongly modified by a weak force at frequency $\omega_p/2$. The effect has generic aspects, which go beyond the model of the parametric oscillator. They manifest most clearly where the decay rate is small, so that the spacing of the intrawell levels of the RWA Hamiltonian significantly exceeds their width. In this regime, a major effect of the coupling to a thermal bath is transitions between the RWA states, see Fig.~\ref{fig:resonance} (a). The transitions are not limited to the neighbouring RWA states even where the relaxation is due to transitions between the neighboring Fock states. Transitions down to the bottom of the well of the RWA Hamiltonian are more likely then toward the barrier top, the rates $W_\downarrow$ are larger than $W_\uparrow$. Therefore the oscillator is mostly localized near one or the other minimum of the well \footnote{In a certain parameter range, along with stable states of parametrically excited vibrations, the quiet oscillator state, which corresponds to the local maximum of the RWA Hamiltonian also becomes stable. In this paper we do not consider this parameter range.}. However, since $W_\uparrow$ is nonzero, the oscillator essentially performs a random walk over the intrawell states. In the course of this walk it can reach the barrier top and then switch to the other well with probability $\sim 1/2$. 

We note the difference between the rates of transitions between the quantum intrawell states sketched in Fig.~\ref{fig:resonance} (a) and the rate of switching between the stable vibrational states. We use the term  ``switching'' to describe transitions between the wells of the RWA Hamiltonian. We find that, in the quantum domain, the change of the switching rates is also exponetial in the amplitude of the force at frequency $\omega_p/2$, with the exponent determined by the ratio of the change of the effective ``potential well'' to the Planck constant.

To set the scene, in Sec.~\ref{sec:Hamiltonian} we introduce the Hamiltonian of the parametrically modulated nonlinear oscillator. We show that the Hamiltonian has two wells and discuss the intrawell dynamics in the presence of a weak force. In Sec.~\ref{sec:quantum_activation} we present the master equation, which describes the effect of the coupling to a thermal bath on the oscillator dynamics. We reduce this equation to the balance equation for the populations of the intrawell states in the weak-damping limit. We explain how this balance equation can be solved within the WKB approximation and how the interwell switching rates are found from this solution. Section~\ref{sec:transition_rates} is technical:  we use the methods of nonlinear dynamics to develop a perturbation theory that allows us to find corrections to the transition rates, which are linear in the amplitude of the extra force. The result is used in Sec.~\ref{sec:LS_finite_nbar} to find the corrections to the intrawell state populations and to the switching rate. The change of the {\it logarithm} of the switching rate is linear in the extra force amplitude. We find the corresponding logarithmic susceptibility and its dependence on $\bar n$ and the oscillator parameters. The explicit expressions are obtained for comparatively high temperatures and close to the bifurcation point where there emerge period-2 vibrations; in particular, we consider the prebifurcation scaling where the motion near the bifurcation point is still underdamped. Section~\ref{sec:conclusion} contains concluding remarks.

The extra force exponentially increases the ratio of the switching rates from the wells it makes shallower and deeper, $W_\mathrm{sw}^\mathrm{shallow}/W_\mathrm{sw}^\mathrm{deep}$. We have $W_\mathrm{sw}^\mathrm{shallow}/W_\mathrm{sw}^{(0)} \approx W_\mathrm{sw}^{(0)}/W_\mathrm{sw}^\mathrm{deep}$, where $W_\mathrm{sw}^{(0)}$ is the switching rate in the absence of the extra force. For concreteness, we calculate the change of the switching rate from the deeper well.

%%%%%%%%%%%%%%%%%%%%%%%%%%%%%%%%%%%%%%%%%%%%%%%%%%%%%%%%%%%%%%%%%%

\section{THE HAMILTONIAN IN THE ROTATING WAVE APPROXIMATION}
\label{sec:Hamiltonian}

If the decay rate of the oscillator $\Gamma$ is small compared to its eigenfrequency $\omega_0$, even a comparatively small periodic modulation of $\omega_0$ at frequency $\omega_p$ close to $2\omega_0$ can lead to bistability. The onset of stable states requires that the oscillator be nonlinear \cite{Landau2004}. Classically, the vibration frequency of a nonlinear oscillator depends on its amplitude. Therefore, as the amplitude of the parametrically excited  vibrations increases, the vibration frequency moves away from $\omega_p/2$, weakening the resonance with the modulation and stabilizing the vibrations.

A simple nonlinearity of the oscillator potential that leads to the stabilization in the lowest order of the perturbation theory is the Duffing (Kerr) nonlinearity. It is relevant to many physical systems and is described by the quartic term in the oscillator coordinate $q_0$. The Hamiltonian of the parametrically modulated Duffing oscillator has the form 
\begin{align}
\label{eq:H_0}
	H^{(0)} &= \frac{1}{2} p_0^2 + \frac{1}{2} q_0^2 \left[\omega_0^2 + F_p\cos(\omega_p t)\right] + \frac{1}{4} \gamma q_0^4,
\end{align}	
where $q_0$ and $p_0$ are the oscillator coordinate and momentum, $F_p$ is the modulation amplitude, $\gamma$ is the nonlinearity parameter, and we have set the oscillator mass equal to unity. 

In the presence of an additional linear force at half the modulation frequency the Hamiltonian becomes 
\begin{align}
\label{eq:full_Hamiltonian}
	H = H^{(0)} - q_0 A_d \cos(\omega_pt/2 + \varphi_d),
\end{align}
where $A_d$ and $\varphi_d$ are the amplitude and phase of the force.

The oscillator dynamics is conveniently described by switching to the rotating frame with the unitary transformation $\hat{U} = \exp[-i \hat{a}^\dagger \hat{a} (\omega_p t/2)]$, where $\hat{a}^\dagger$ and $\hat a$ are the ladder operators,  $\hat{a}=(\hbar \omega_p)^{-1/2} (i \hat p_0+\omega_p \hat q_0/2 )$. In the rotating wave approximation (RWA) the von Neumann equation for the oscillator density matrix $\hat \rho$ in the rotating frame reads 
\begin{align}
\label{eq:master_eq_no_damping}
	d\hat \rho/d\tau = i \lambda^{-1} [\hat \rho,\hat g], \qquad \tau = tF/2\omega_p.
\end{align}	
%	%
Here $\hat g$ is the scaled RWA Hamiltonian, 
\begin{align}
\label{eq:full_g}
&\hat g\equiv  \hat g(Q,P) =\hat g^{(0)}+ \alpha_d \hat g^{(1)}, \nonumber\\
&\hat g^{(0)} = \frac{1}{4} (Q^2+P^2)^2 + \frac{1}{2} (1-\mu) P^2 - \frac{1}{2} (1+\mu) Q^2\,,\nonumber\\
&\hat g^{(1)}= - \left[P \cos \varphi_d + Q \sin \varphi_d\right].
\end{align}
In Eqs.~(\ref{eq:master_eq_no_damping}) and (\ref{eq:full_g}) $Q$ and $P$ are (the operators of) the dimensionless coordinate and momentum, $\tau = t F/2\omega_p$ is the dimensionless time, and $\alpha_d = A_d \sqrt{6\gamma/F_p^3}$ is the scaled amplitude of the extra additive force. Here and in what follows we use the superscripts $0$ and $1$ to indicate the parameters of the oscillator unperturbed by the extra force and the perturbation, respectively. For a weak extra force that we consider $\alpha_d \ll 1$.

In terms of the ladder operators in the rotating frame
\begin{align}
\label{eq:in_terms_of_ladder}
	\hat Q=&i(\lambda/2)^{1/2}(\hat a - \hat a^\dagger), \quad \hat P=(\lambda/2)^{1/2}(\hat a + \hat a^\dagger),\nonumber\\
	\hat g=&\lambda^2 (\hat a^\dagger \hat a + 1/2)^2 + \frac{\lambda}{2} (\hat a^2 + \hat a^{\dagger^2})   - \mu \lambda \hat a^\dagger \hat a \nonumber\\
	&- \alpha_d \sqrt{\frac{\lambda}{2}} \left(\hat a e^{i\varphi_d} + \mathrm{c.c.}\right)\,,
\end{align}
where $\lambda = 3\gamma \hbar/F_p \omega_p$ is the scaled Planck constant, $[\hat Q,\hat P]=i\lambda$. In the absence of the extra force the Hamiltonian $\hat g(Q,P)$ depends only on one parameter, the scaled detuning 
\begin{align}
\label{eq:mu}
\mu = 2\omega_p \delta \omega /F_p, \qquad \delta \omega = \omega_p/2 - \omega_0.
\end{align}

We note that, since we switched to the rotating frame at half the modulation frequency, in the absence of extra additive force $\propto A_d$  the Hamiltonian $\hat g\0(Q,P)$ is not the Floquet (quasienergy) Hamiltonian. To avoid confusion we call $\hat{g}$ the RWA Hamiltonian, and its eigenvalues the RWA energy values.

%%%%%%%%%%%%%%%%%%%%%%%%%%%%%%%%%%%%%%%%

\subsection{Intrawell dynamics}
\label{subsec:intrawell_dynamics}

The unperturbed RWA Hamiltonian $\hat g\0(Q,P)$ is a symmetric function of $Q$ and $P$. For $-1<\mu<1$ it has two minima. They lie on the $Q$-axis at ${Q^{(0)}_{\pm}=\pm \sqrt{1+\mu}}$. At the minima  ${g^{(0)}(Q^{(0)}_\pm,P=0) \equiv g_{\mathrm{min}}^{(0)} = -(1+\mu)^2/4}$.  In the laboratory frame, the minima correspond to parametrically excited vibrations with opposite phases; the coordinate $q_0(t)$ is $\propto -Q_\pm^{(0)}\sin(\omega_pt/2)$.

The minima are separated by a saddle point at $Q=P=0$. Classical Hamiltonian dynamics inside the symmetric wells of the function $g\0(Q,P)$ is well understood \cite{Marthaler2006}. The oscillator moves along closed intrawell trajectories with constant RWA-energy, $g\0(Q,P)=g$. The trajectories in the opposite wells are mirror-symmetric and, for a given $g$, have the same frequency $\omega\0(g)$.  

We now consider the effect of  the force  $\propto \alpha_d$ on the classical trajectories inside the wells. The force breaks the symmetry of $g(Q,P)$, as it tilts it. The direction of the tilt is determined by the phase $\varphi_d$. For a weak force, $\alpha_d \ll 1$, the function $g(Q,P)$ still has two wells, which are now asymmetric and may have different depths. For the phase $\varphi_d = (2k+1) \pi/2$ with integer $k$ the tilt is along the $Q$ axis. The minimum of one well shifts towards the origin and the well depth decreases, while the other well shifts away from the origin and its depth increases. For $\varphi_d =k\pi$ the wells are shifted along the $P$ axis. In the general case, to the first order in $\alpha_d$ the values of $g(Q,P)$  at the minima are
\begin{align}
\label{eq:minima}
		&g_{\mathrm{min}{}} = g_{\mathrm{min}}^{(0)} \pm  \alpha_d g_{\mathrm{min}}^{(1)}\,, \quad g_{\mathrm{min}}^{(1)} = -\sqrt{1+\mu} \sin \varphi_d\,.
\end{align}
where the signs ``$+$'' and ``$-$'' refer to the wells at $Q>0$ and $Q<0$, respectively. 
The saddle point of $g(Q,P)$ shifts from $Q=P=0$ to $Q=-\alpha_d\sin\varphi_d/(1+\mu)$ and $P=-\alpha_d\cos\varphi_d/(1-\mu)$. The value $g_\mathrm{saddle}$ of $g(Q,P)$ at the saddle point does not change, to the first order in $\alpha_d$.

\begin{figure}[H]
	\centering
	\includegraphics[width=1\linewidth]{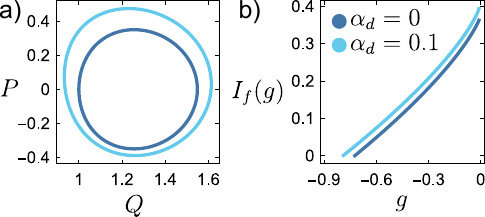}
	\caption{(a) Classical Hamiltonian trajectories $g=~$const. inside the well of the RWA Hamiltonian $g(Q,P)$ at $Q>0$ for $g=-0.6$. (b) The dependence of the action variable on the RWA energy $g$. Both panels refer to $\mu=0.7$ and $\varphi_d = \pi/6$.}
	\label{fig:trajectories_action}
\end{figure}

The change of $g(Q,P)$ due to the force $\propto \alpha_d$  leads to a change of the Hamiltonian intrawell trajectories. Generally, we expect the frequency $\omega(g)$ to change, too. The frequency can be found by calculating the action variable $I_f$ as a function of the RWA energy $g$, 
\begin{align*}
	\omega^{-1}(g) = \pdv{I_f(g)}{g}\,, \qquad I_f(g) = \frac{1}{2\pi} \oint P(Q|g) d{}{}{Q}\,,
\end{align*}
where the integral is taken over the trajectory with a given $g$ inside the well; $P(Q|g)$ is given by the equation $g(Q,P)=g$. We use the subscript $f$ to indicate that $I_f$  refers to the full Hamiltonian function $g=g^{(0)} + \alpha_d g^{(1)}$.  In Appendix~\ref{sec:app_classical_motion} we show that, to the first order in $\alpha_d$, the actions in the two wells change by $\pm \alpha_d I_f^{(1)} = \pm \alpha_d \sin (\varphi_d)/2$. Remarkably this change is independent of $g$, see Fig. \ref{fig:trajectories_action} (b). Then, to the first order in $\alpha_d$, the frequency of intrawell classical motion is not changed by the linear force. This has interesting consequences for the energy spectrum of the RWA Hamiltonian.

%

%%%%%%%%%%%%%%%%%%%%%%%%%%%%

\subsection{The RWA energy levels}
\label{subsec:energy_levels}

For the dimensionless RWA Hamiltonian $\hat g$, the distance between the eigenvalues, i.e., between the RWA energy levels, is proportional to the dimensionless effective Planck constant $\lambda$. As indicated earlier, of interest for quantum information and for many other physics problems is the case where there are many quantum states $\ket{n}$ inside the wells of the function $g(Q,P)$. This implies that $\lambda\ll 1$. Moreover, we will be interested in the regime where the extra force, although weak, is still ``quantum strong'', that is the force-induced shift of the RWA energy levels is larger than the level spacing, which implies that $\lambda\ll \alpha_d$. Respectively, the shift of the minima $| \alpha_d g_{\min}^{(1)}|$ significantly exceeds the spacing of the levels as well.

Since for $\alpha_d = 0$ the wells of $g(Q,P)$ are symmetric and intrawell states of different wells are in resonance, the eigenstates of $\hat g$ are given by the tunnel split symmetric and antisymmetric combinations of the intrawell states. As $\alpha_d$ increases, the levels in different wells shift away from each other and resonant tunneling is suppressed. The eigenstates $\ket{n}$ of $\hat g$ are well-localized intrawell states,
\[\hat g\ket{n} = g_n\ket{n}.\]
where $g_n$ are the intawell RWA energies.

We note that a part of the states in different wells may become resonant again for certain values of $\alpha_d$. In the semiclassical limit, the distance between the intrawell levels $g_n$ is $\lambda\omega(g)$. Therefore, given that $\omega(g)$ is not changed by the force, for such $\alpha_d$, simultaneously, all levels in the shallow well come to resonance with the levels in the deeper well. 

%%%%%%%%%%%%%%%%%%%%%%%%%%%%%%%%%%%%%

\section{QUANTUM ACTIVATION}
\label{sec:quantum_activation}

Coupling the oscillator to a thermal bath leads to dissipation. In the absence of modulation, dissipation is associated with transitions between the Fock states of the oscillator (i.e., the eigenstates of the Hamiltonian $H^{(0)}$ for $F_p=0$), which are accompanied by emission and absorption of excitations of the bath. A major dissipation process is associated with transitions between neighbouring Fock states, with energy exchange $\approx \hbar\omega_0$. Classically, it leads to a viscous-type friction force $-2\Gamma\dot q_0$. We will assume that the coupling is weak, so that the friction coefficient $\Gamma$ is small compared to the oscillator eigenfrequency $\omega_0$. If certain well-understood conditions are met, the oscillator dynamics is Markovian on the time scale slow compared to $\omega_0^{-1}$ \cite{Bachtold2022a}. 

Resonant parametric modulation does not open new dissipation channels, to the leading order in $F_p$.  However, now the state nomenclature is changed: dissipative transitions between the Fock states of the oscillator translate to transitions between the RWA states, since the latter states are linear combinations of the former states. Because the overlapping of the RWA states in different wells is exponentially small for small $\lambda$, of primary importance are transitions between the states within the same well. It is characteristic that the transitions between neighboring Fock states are projected onto transitions between not only neighbouring, but also remote RWA states. In the absence of an extra force, the rates $W_{nn'}$ of transitions between intrawell states $\ket{n}\to \ket{n'}$ were calculated earlier \cite{Marthaler2006}.   

If the rates $W_{nn'}$ are small compared to the levels spacing $\lambda\omega(g)$, the dynamics of the parametric oscillator in slow time $\tau$, Eq.~(\ref{eq:master_eq_no_damping}), can be described by the balance equation for the intrawell state populations $\rho_n=\bra{n}\rho\ket{n}$,  
\begin{align}
\label{eq:balance}
	\pdv{\rho_n}{\tau} &= -\sum_{n'} (W_{n n'}\rho_n - W_{n' n} \rho_{n'})\,,\nonumber \\
	W_{nn'} &= W_{nn'}^{\mathrm{(e)}} + W_{nn'}^{\mathrm{(abs)}}\,, \quad  W_{n n'}^{\mathrm{(abs)}} = 2\kappa  \bar{n} \abs{\Braket{n|\hat{a}|n'}}^2\,, \nonumber \\
	W_{n n'} ^{\mathrm{(e)}} &= 2\kappa(\bar{n}+1) \abs{\Braket{n|\hat{a}^\dagger|n'}}^2  \,.
\end{align}
Here, $\bar{n} = [\exp{\hbar \omega_p/2k_BT} -1]^{-1}$ is the oscillator thermal occupation number; $\kappa$ is the scaled friction coefficient, $\kappa = 2\Gamma\omega_p/F_p$. We omitted the indices that label the wells. 

In the semiclassical approximation, where $n,n'\gg 1$ and $|n-n'|$ is not too large, we can express the matrix elements in Eq.~(\ref{eq:balance}) by the Fourier transforms of the complex amplitudes of intrawell vibrations \cite{Landau2004a}, 
\begin{align}
\label{eq:semiclassical_W}
&\braket{n+m|\hat a|n}\approx a_m(g_n),\nonumber\\
&a_{m}(g) = \frac{1}{2\pi} \int_{0}^{2\pi} d{}{}\phi\, a(\tau,g) e^{-im\phi}\,,
\end{align}
where $\phi = \omega(g) \tau$ and $a(\tau,g) =(2\lambda)^{-1/2} [P(\tau,g) - i Q(\tau,g)]$; $Q(\tau,g)$ and $P(\tau,g)$ are the solution of the Hamiltonian equations of motion for the Hamiltonian in Eq.~(\ref{eq:full_g}). The  expressions for the transition rates in terms of $a_m(g)$ have the form
\begin{align}
\label{eq:transition_rates_Fourier}
	W_{n+m\, n}^{\mathrm{(e)}} &= 2\kappa (\bar{n}+1) \abs{a_{-m}(g_n)}^2 \,, \nonumber \\
	W_{n+m\, n}^{\mathrm{(abs)}} &= 2\kappa \bar{n} \abs{a_{m}(g_n)}^2 \,. %\nonumber
\end{align}

An explicit calculation of the matrix elements $\braket{n|\hat a|n'}$ shows \cite{Marthaler2006} that, in the absence of the extra force, the transition rates $W_{nn'}$ satisfy the condition $W_{nn'}>W_{n'n}$ for $n>n'$, if we use the convention that the states $\ket{n}$ are counted off from the bottom of the well of $g(Q,P)$. This strong inequality cannot be broken by a weak extra force. It means that the oscillator is more likely to go down towards the bottom of the well than going up away from it. This corresponds to relaxation to a classically stable state, in quantum terms.

The transition probabilities have two contributions corresponding to absorption and emission of excitations of the heat bath. Importantly, even in the regime where the thermal occupation number $\bar{n}$ can be assumed to be zero and only emission processes are relevant, transitions away from the bottom of the well still have nonzero probability, populating excited intrawell states. This effect was termed quantum heating \cite{Dykman2011} and was directly observed in the experiment \cite{Ong2013}. In the random walk over intrawell states, once the oscillator makes a transition away from the bottom of the well, it is more likely to go back down, but it still can go further up. Ultimately, if it reaches the top of the well, where $g(Q,P)=0$, it can switch to another well. Such switching is similar to thermal activation in equilibrium systems and has been called quantum activation. 

%%%%%%%%%%%%%%%%%%%%%%%%%%%%%%%%%%%%%%%%%%%%%%

\subsection{Discrete WKB approximation}
\label{subsec:discrete_WKB}

In order to calculate the  rate $W_\mathrm{sw}$ of interwell switching we investigate the quasistationary distribution over intrawell states described by Eq.~(\ref{eq:balance}) in which we set $\partial\rho_n/\partial \tau = 0$. Such approach is justified by the strong inequality $W_\mathrm{sw}\ll \kappa$ and is similar to the analysis of the switching rate in thermal equilibrium systems \cite{Kramers1940}. To find the quasistationary distribution in the semiclassical range, where the number of intrawell states is large, we use the ansatz 
\begin{align}
\label{eq:quasistationary}
\rho_n = \exp[-R(g_n)/ \lambda]
\end{align}
and use that (i) $R(g)$ is a smooth function of $g$ and that (ii) $W_{n+m\,n} \approx W_{n\,n-m}$. The latter condition is based on the fact that the rates $W_{n\,n+m}$ fall off exponentially with the increasing $|m|$ and that the typical $n$ are much larger than the typical $|m|$. Then Eq.~(\ref{eq:balance}) is reduced to a set of equations for $R'(g)$,
\begin{align}
\label{eq:xi_stationary}
	&\sum_m W_{n+m\,n} (1- \xi_n^m) = 0\,, \\
	&\xi_n = \exp[-\omega(g_n) R'(g_n)]\,, \qquad R'(g) = dR(g)/dg\,,\nonumber
\end{align}
where we used $g_{n+m} - g_n \approx m \lambda\omega(g_n)$. The quasistationary distribution (\ref{eq:quasistationary}) inside a well is determined by the function
\begin{align*}
	R(g) &= \int_{g_{\mathrm{min}}}^g R'(x)\, dx\,.
\end{align*}

The rate of switching from a well $W_\mathrm{sw}$ is approximately given by the probability per unit time to reach a state with $g$ close to the saddle-point energy $g_\mathrm{saddle}$,
\begin{align}
\label{eq:R_A_defined}
	W_{\mathrm{sw}} &= C_{\mathrm{sw}} \times \exp(-R_\mathrm{A}/\lambda)\,, \quad R_\mathrm{A} = R(g_{\mathrm{saddle}})\,.
\end{align} 
To the first order in the amplitude of the extra force, $g_\mathrm{saddle}=0$. The prefactor $C_{\mathrm{sw}}$ in Eq.~(\ref{eq:R_A_defined}) is proportional to the decay rate in the absence of modulation $\kappa$. 

%%%%%%%%%%%%%%%%%%%%%%%%%%%%%%%%%%%%%%%%%%%%%%%%%%%%%%%%%%%%%%%%%%%%%%%%%%%%%%%%%%%%%

\section{Effect of the extra force on the transition rates}
\label{sec:transition_rates}

We emphasize again that we consider the dynamics in one of the wells of $g(Q,P)$. The extra force not only shifts the RWA energies of the intrawell states, but also modifies the transitions rates $W_{n\,n + m}$ by changing the matrix elements $a_{m}(g_n) = \braket{n+m|\hat{a}|n}$.  We calculate the corrections to $a_m(g)$ in Eq.~(\ref{eq:semiclassical_W}) assuming that the perturbation is classically weak, $\alpha_d \ll 1$. 

Where there is no extra force, the classical trajectories $Q(\tau,g), P(\tau,g)$ can be expressed in terms of the Jacobi elliptic functions, leading to simple expressions for their Fourier components \cite{Marthaler2006},  see Appendix \ref{sec:app_classical_motion}. The force changes the trajectories, and we have not found analytical expressions for them. Examples of the trajectories with and without a weak extra force are shown in Fig. \ref{fig:trajectories_action}~(a).

%%%%%%%%%%%%%%%%%%%%%%%%%%%%%%%%%%%%%%%%%%%%%%%%%%%%%%%%%%%%%%%

\subsection{Action-angle variables}
\label{subsec:action_angle}

It is convenient to find the force-induced corrections to $a_m(g)$ by switching to the action-angle variables $I,\psi$ of the unperturbed system. Formally, we proceed by considering a system with the coordinate $q$, momentum $p$, and the Hamiltonian function $g^{(0)}(q,p)$ and make a standard canonical transformation 
\begin{align}
\label{eq:transformation}
&p=\partial S(q,I)/\partial q, \quad \psi = \partial S(q,I)/\partial I,\nonumber\\
&I(g) = \frac{1}{2\pi} \oint p(q|g) dq\,, \quad g^{(0)}(q,p)=g, \nonumber\\
&q(I;\psi+2\pi) = q(I;\psi), \quad p(I;\psi+2\pi) = p(I;\psi),
\end{align}
where the generating function $S$ is the action calculated for the unperturbed Hamiltonian $g^{(0)}$ and $p(q|g)$ is the momentum calculated from the equation $g^{(0)}(q,p)=g$. The explicit relation between $(q,p)$ and $(I,\psi)$ can be found from the Fourier components of $q,p$ calculated for the Hamiltonian $g^{(0)}$, see Appendix~\ref{sec:app_classical_motion}. The function $I(g)$ satisfies the equation
\[d{}I/d{}g = 1/\omega^{(0)}(g), \quad g= g^{(0)}\bigl(q(I;\psi),p(I;\psi)\bigr)=g^{(0)}(I),\]
where $\omega^{(0)}(g)$ is the frequency of intrawell vibrations in the absence of the extra force; it is given in  Appendix~\ref{sec:app_classical_motion}. The above equation gives the Hamiltonian function $g\0$ as a function of the action, $g^{(0)}(I;\psi)\equiv g^{(0)}(I)$.

In what follows we will consider the coordinate and momentum of the oscillator $Q,P$ in the presence of the extra force as functions of $I,\psi$. We define their functional form as 
\begin{align*}
&Q(I;\psi) = q(I;\psi),\quad P(I,\psi) = p(I;\psi); \\
& g^{(0)}(Q,P) = g^{(0)}(I).
\end{align*}
We distinguish between the functions $Q(\tau, g)$ and $Q(I;\psi)$ by making use of the semicolon. The same convention is used for $P(I;\psi)$. 

The full RWA Hamiltonian is time-independent. However, since $I$ and $\psi$ are defined with respect to $g^{(0)}$, in the presence of the extra force the action $I$ depends on time and the time dependence of $\psi$ is changed compared to the case where there is no extra force. 
To find the time dependence of $I,\psi$ one has to express the Hamiltonian function $g = g^{(0)} + \alpha_d g^{(1)}$ in terms of $I,\psi$. To emphasize this form of the Hamiltonian we write it as $G(I;\psi)=G^{(0)}(I;\psi) + \alpha_d G^{(1)}(I;\psi)$, where 
\[G^{(\alpha)}(I;\psi) = g^{(\alpha)}\bigl(Q(I;\psi),P(I;\psi)\bigr), \quad \alpha=0,1.\]
Since $G^{(0)}$ and $G^{(1)}$ are obtained from Eq.~(\ref{eq:full_g}) by substituting $Q= Q(I;\psi),\, P=P(I;\psi)$, they are periodic in $\psi$. The equations of motion for $I,\psi$ read
\begin{align}
\label{eq:action_angle}
	&\frac{dI}{d\tau} = -\pdv{G}{\psi}\,, \qquad \frac{d\psi}{d\tau} = \pdv{G}{I}\,,\nonumber\\
	 &G\equiv G(I;\psi) = G^{(0)} + \alpha_d G^{(1)}= G(I;\psi+2\pi)\,.
\end{align}
These are Hamiltonian equations describing trajectories with a constant RWA energy $G(I;\psi)=g$. By  construction 
\begin{align}
\label{eq:G_0}
&G^{(0)}(I;\psi) = \int_0^I dI'\,\Omega^{(0)}(I')+g_\mathrm{min}^{(0)}\,, \nonumber\\ 
&\Omega^{(0)}(I) \equiv \omega^{(0)}\bigl(g^{(0)}(I)\bigr)\,.
\end{align}
The frequency $\Omega^{(0)}(I)$ is the vibration frequency for the unperturbed system as a function of the action $I$.

To find the extra-force induced corrections to the Fourier components $a_m(g)$, we will seek corrections to $I$ and $\psi$ for a given RWA energy $g= G^{(0)}+ \alpha_d G^{(1)}$. Because ultimately we need corrections to the Fourier components of the variables $Q,P$, the analysis is slightly different from the conventional analysis of nonlinear dynamics \cite{Arnold1989}. 

From Eq. (\ref{eq:action_angle}),  to the first order in $\alpha_d$
\begin{align}
\label{eq:first_order_I_psi}
	I(\tau) &= I^{(0)} + \alpha_d I^{(1)}(\tau), \nonumber \\
\psi(\tau) &= \Omega^{(0)}(\bar I)\tau + \alpha_d \psi^{(1)} (\tau)\,.
\end{align}
Here and in what follows the overline means period averaging. Both $I^{(1)}$ and $\psi^{(1)}$ are periodic functions of time, as they are determined by  $G^{(1)}$. In particular, $\psi\1(\tau)$ comes from integrating over time the term $\partial G\1/\partial I$ in Eq.~(\ref{eq:action_angle}) for $d\psi/d\tau$, but this is not the only first-order correction to $\psi(\tau)$, as explained below.

%%%%%%%%%%%%%%%%%%%%%%%%%%%%%%%

\subsection{Vibration frequency for a given intrawell energy}
\label{subsec:vibration_frequency}

The vibration frequency inside a well of $g(Q,P)$ is determined by the secular term $\propto \tau$ in $\psi(\tau)$. There are two extra-force induced contributions to this term. One comes from the difference between $\bar I$ and $I\0$ in the term  $\propto \Omega^{(0)}(\bar I)$ in Eq,~(\ref{eq:first_order_I_psi}). To the first order in $\alpha_d$, the value of $\bar I$ for a given $g$ has to be found from the equation $G^{(0)}(\bar I) + \alpha_d \bar G^{(1)}=g$. From this equation we find $\bar I = I^{(0)} + \alpha_d\bar I^{(1)}$ with $\bar I^{(1)} = -\bar G^{(1)}/\Omega^{(0)}$ (both $\bar G^{(1)}$ and $\Omega^{(0)}$ here are calculated for $I=I^{(0)} $). 

The second secular contribution is contained in $\psi\1(\tau)$ and comes from the term $\partial \bar G^{(1)}/\partial I$ in Eq.~(\ref{eq:action_angle}) for $d\psi/d\tau$. From Eqs.~(\ref{eq:full_action_correction}) and (\ref{eq:first_order_action}), $\bar G^{(1)} =\bar C \Omega^{(0)}(I^{(0)})$, where $\bar C$ is independent of $I$. Therefore the secular term in $\psi^{(1)}$ cancels the secular correction $\propto  \bar I^{(1)}$ in $\psi(\tau)$. Indeed, $(d\Omega^{(0)}/dI) \bar{I}^{(1)} + (d\bar G^{(1)}/dI) = 0$. As a result, to the first order in $\alpha_d$,  the secular term in $\psi(\tau)$ is $\omega\0(g)\tau$, and therefore the oscillating terms are $\propto \exp[in\omega\0(g)\tau]$ with integer $|n|>0$,  
\begin{align}
\label{eq:psi_1_general}
\psi(\tau) = \omega^{(0)}(g)\tau + \alpha_d \sum_{n\neq 0} \psi_n^{(1)}\exp[in\omega^{(0)}(g)\tau]\,,
\end{align}
where $\psi_n\1\equiv \psi_n\1(g)$ are the Fourier components of $\psi^{(1)}$. To the same order of the perturbation theory, $I^{(1)}(\tau)$ is a sum of terms $\propto\exp[in\omega^{(0)}(g)\tau]$ with $n \neq 0$.  They, as well as $\psi_n\1$, are immediately expressed in terms of the Fourier components of $G^{(1)}$, see Appendix~\ref{sec:corrections_to_a}.

We can now calculate the corrections to the Fourier components $a_m(g)$ to the first order in $\alpha_d$. We write 
\[a(\tau,g) \equiv a(I;\psi)= [P(I;\psi)-iQ(I;\psi)]/\sqrt{2\lambda}.\] 
The functions $Q(I;\psi)$ and $P(I;\psi)$ are periodic functions of $\psi$, while $I\equiv I(\tau)$ and $\psi\equiv \psi(\tau)$ are periodic functions of time with frequency $\omega\0(g)$. The Fourier components 
\[a_m(g)=\frac{1}{2\pi}\int_0^{2\pi}d\phi\,a(\tau,g)\exp(-im\phi)\]
with $\phi = \omega\0(g)\tau$ are determined  by the Fourier components of $a(I;\psi)$. To find $a_m(g)$ we expand $a(I;\psi)$ to the first order in $I^{(1)}, \psi^{(1)}$, and  use the Fourier series for $I^{(1)}, \psi^{(1)}$, cf. Eq.~(\ref{eq:psi_1_general}).  This gives 
\begin{align}
\label{eq:Fourier_a_general}
	& a_{m}(g) \approx a_{m}^{(0)}(g) + \alpha_d a_{m}^{(1)}(g)\,,\nonumber\\
	&a_m^{(1)} = \sum_k\left[J_{mk} a_{m-k}\0 + L_{mk}(d{}a_{m-k}\0/d{}g) \right]\,.
\end{align}
Explicit expressions for the parameters $J_{mk}\equiv J_{mk}(g)$ and $L_{mk}\equiv L_{mk}(g)$  follow from Eq.(\ref{eq:a_m_1}) in Appendix~\ref{sec:corrections_to_a}. They apply in the range $g>\max \{g_{\mathrm{min}} , g_{\mathrm{min}}^{(0)}\}$. It is straightforward also to find higher-order corrections to $I,\psi$ and then to $a(I;\psi)$. We note that the vibration frequency will be shifted from $\omega^{(0)}(g)$ in the second order in $\alpha_d$.

\subsection{Transition rates}
\label{subsec:first_order_rates}

Corrections to the emission and absorption transition rates $W_{nn'}^\mathrm{(e)}$ and $W_{nn'}^\mathrm{(abs)}$ are found by inserting the expansion for $a_{m}(g_n)$ into Eq.~(\ref{eq:transition_rates_Fourier}). This gives the rates in the form of perturbation series
\begin{align}
\label{eq:rates_corrections}
	&W_{n+m\, n} = W_{n+m\, n }^{(0)} + \alpha_d W_{n+m\, n}^{(1)} + \ldots \,, \nonumber\\
	&(W_{n+m\, n}^\mathrm{(e)})^{(1)} = 4\kappa(\bar n +1)\,\mathrm{Re}\,\left[a_{-m}^{(0)}{}^*(g_n)a_{-m}^{(1)}(g_n)\right]\nonumber\\
&(W_{n+m\,n}^\mathrm{(abs)})^{(1)} = 4\kappa \bar n\,\mathrm{Re}\,\left[a_{m}^{(0)}{}^*(g_n)a_{m}^{(1)}(g_n)\right]	
\end{align}
In turn, this allows finding  corrections to the populations of the intrawell states $\rho_n$ and the rate $W_\mathrm{sw}$ of interwell switching. 

An important feature of the rates $W^{(1)}$ is their specific dependence on the phase of the extra force. We find that $W^{(1)}_{n+m\,n} \propto \sin\varphi_d$. This is in spite of $a_m^{(1)}(g)$ being a linear combination of $\cos\varphi_d$ and $\sin\varphi_d$. Formally, this is a consequence of $a_m^{(0)}$ being purely imaginary, see Eq.~(\ref{eq:fourier_unperturbed}), whereas the term $\propto \cos\varphi_d$ in $a_m^{(1)}$ is real, as shown in Appendix~\ref{sec:corrections_to_a}, see also  Appendix~\ref{sec:q_perturbation_theory}, and drops out from $\mathrm{Re}\,[(a_m\0)^*\, a_m^{(1)}]$. We note that the fact that $\cos\varphi_d$ does not affect the quantum dynamics is a consequence of the approximation of slow relaxation, where the dynamics is described by the balance equation for the state populations (\ref{eq:balance}).

%%%%%%%%%%%%%%%%%%%%%%%%%%%%%%%%%%%%%%%%%%%%%%%%%%%%%%%%%%%%%%%%%%%%%%%%%%%%%%%%%

\section{LOGARITHMIC SUSCEPTIBILITY}
\label{sec:LS_finite_nbar}

The number of intrawell states of the oscillator is $\sim 1/\lambda \gg 1$. Therefore corrections to the rates of interstate transitions accumulate to exponentially large changes of the populations of highly excited states and  to  a change of the quantum activation energy $R_\mathrm{A}$ of the interwell switching. In particular, $R_\mathrm{A}$ acquires a linear in $\alpha_d$ correction, so that the switching rate $W_\mathrm{sw} \propto \exp(-R_\mathrm{A}/\lambda)$ changes exponentially strongly for $\alpha_d\gg \lambda$ even where the force is weak, $\alpha_d\ll 1$. The factor multiplying  $A_d\propto \alpha_d$ in the expression for $R_\mathrm{A} \propto \lambda|\log W_\mathrm{sw}|$ is the logarithmic susceptibility (the concept of  the logarithmic susceptibility applies also in a more general case \cite{Smelyanskiy1997b,Ryvkine2006a}, in particular where the frequency of the extra force differs from $\omega_p/2$).

In the absence of an extra force the oscillator displays qualitatively different dynamics depending on the thermal occupation number $\bar n$ \cite{Marthaler2006}. For $\bar n=0$ the rates of  transitions between the states  are determined by emission of excitations of the medium,  $W_{n+m\,n}=W_{n+m\, n}^\mathrm{(e)}$, and the oscillator has detailed balance. In contrast, interstate transitions due to absorption of excitations of the medium, whose rates are $\propto \bar n$, break the detailed balance. Concurrently, beyond a narrow range of $\bar n$ that goes to zero as $\lambda\to 0$, the occupation of highly excited intrawell states is exponentially increased due to the absorption-induced transitions. We focus on the regime where detailed balance is broken due to a finite $\bar{n}$.

Direct perturbation theory allows us to find the driving-induced corrections to the function $R(g)$ that gives the quasistationary intrawell probability distribution (\ref{eq:quasistationary}). Plugging Eq.~(\ref{eq:rates_corrections}) into Eq.~(\ref{eq:xi_stationary}) gives the derivative $R'(g)$ to the first order in $\alpha_d$,
\begin{gather}
	R'(g) \approx R'^{(0)}(g) + \alpha_d R'^{(1)}(g) \,, \quad	R'^{(0)}(g) = - \frac{\log(\xi^{(0)})}{\omega(g)}\,, \nonumber \\
	R'^{(1)}(g_n) = -\frac{\sum_m W_{n+m \,n}^{(1)} (1-(\xi^{(0)})^m)}{\omega(g) \sum_m m W_{n+m \, n}^{(0)} (\xi^{(0)})^{m}}\,. \label{eq:correction_rprime}
\end{gather}
Here $\xi^{(0)}\equiv \xi^{(0)}(g)=\exp[-\omega(g)R'^{(0)}(g)]$ is the solution of Eq. (\ref{eq:xi_stationary}) in the absence of an extra force. Equation (\ref{eq:correction_rprime}) applies if $g$ is larger than $g_\mathrm{min}^{(0)}$ and the corrected value $g_\mathrm{min}$, Eq.~(\ref{eq:minima}), in the considered well. If $g_\mathrm{min} < g_\mathrm{min}^{(0)}$, $R'(g)$ in the range $g\in (g_{\mathrm{min}{}}, g_\mathrm{min}^{(0)})$ can be found by noting that the motion of the oscillator is harmonic vibrations about the minimum of $g(Q,P)$.  

The correction to the function $R(g)$ is  found by integrating $R'(g)$ over $g$ inside the well, with the boundary condition $R(g_{\mathrm{min}{}})=0$. There are two regions of integration: from $g_\mathrm{min}^{(0)}$ to $g=0$ and from $g_{\mathrm{min}{}} $ to $g\0_\mathrm{min}$. In the first region $R'(g)=R'{}\0 + \alpha_d R'{}\1$, to the first order in $\alpha_d$. The second region is a narrow range with width $ \alpha_d |g_\mathrm{min}^{(1)}|$, and here one can disregard the term $R'^{(1)}$ and use for $R'(g)$ its value $R'{}\0_{\mathrm{min}}\equiv R'{}\0(g\0_\mathrm{min})$ at the bottom of the unperturbed well of $g(Q,P)$,
\begin{align}
\label{eq:Rmin}
	R'{}\0_{\mathrm{min}} &= \frac{1}{2} (1+\mu)^{-1/2} \log(\frac{(\mu+2) (2\bar n +1) + 2\sqrt{1+\mu}}{(\mu +2)(2 \bar n +1) - 2\sqrt{1+\mu}})\,
\end{align}
(cf. \cite{Marthaler2006}). Then in the whole range $g>\max\{g_\mathrm{min}, g_{\mathrm{min}}^{(0)}\}$ 
\begin{align}
\label{eq:R_correction}
	&R(g) \approx R^{(0)}(g) + \alpha_d R^{(1)}(g)\,, 
	\quad R^{(0)}(g) = \int_{g_{\mathrm{min}}^{(0)}}^{g} R'^{(0)}(g), \nonumber\\
	& R^{(1)}(g) = \int_{g_{\mathrm{min}}^{(0)}}^{g} R'^{(1)}(g) d{} g +\alpha_d^{-1} (g_{\mathrm{min}}^{(0)} - g_{\mathrm{min}{}}) R'^{(0)}_{\mathrm{min}}\,. 
\end{align}
Since the corrections $W_{n+m \, n}^{(1)}$ and the shift of the well minimum $ \pm \alpha_d g_{\mathrm{min}}^{(1)}$ are proportional to $\sin \varphi_d$, the change $R^{(1)}(g)$ is also proportional to $\sin \varphi_d$.

The correction to the activation energy $R_\mathrm{A}$ is given by 
\begin{align}
	\alpha_d R_\mathrm{A}^{(1)} = \alpha_d R^{(1)}(g=0),  \label{eq:def:R_A}
\end{align}
where we used that, to the first order in the linear force, the saddle point of $g(Q,P)$ remains at $g=0$. As a result, the switching rate has an additional exponential factor
\begin{align}
\label{eq:R_!_defined}
	W_{\mathrm{sw}} &= W_{\mathrm{sw}}^{(0)} \times \exp(-\alpha_d R_\mathrm{A}^{(1)}/ \lambda)\,.
\end{align}
Here $W_{\mathrm{sw}}^{(0)}$ is the switching rate in the absence of the extra force. The exponent in the ratio $W_{\mathrm{sw}}/W_{\mathrm{sw}}^{(0)}$  is proportional to the ratio $\alpha_d/\lambda$. For $\alpha_d \gg \lambda$ a weak extra force leads to an exponentially strong change of the switching rate, with the exponent linear in the force amplitude. The change of the switching rate is thus described by the logarithmic susceptibility, which is given by $(\alpha_d/A_d) R_\mathrm{A}^{(1)}$ and is independent of the amplitude $A_d$ of the extra force.

One can see that $W_{n+m \, n}^{(1)}$, and thus $R'{}\1$, have opposite signs in the wells of $g(Q,P)$ with $Q>0$ and $Q<0$. Therefore the quantum activation energy  $R_\mathrm{A}^{(1)}\propto \sin\varphi_d$ also has opposite signs for  these wells, i.e., for the parametrically excited vibrations with the coordinate in the laboratory frame $q_0(t)\propto -\sin(\omega_pt/2)$  and $q_0(t)\propto \sin(\omega_pt/2)$, respectively. The difference of the switching rates from different wells is most pronounced for the phase of the extra field $\varphi_d =\pm\pi/2$. The same dependence on $\varphi_d$ holds in the classical regime \cite{Ryvkine2006a}.

For $0 <\varphi_d < \pi$ the well at $Q>0$ is the deeper well of $g(Q,P)$, as seen from Eq.~(\ref{eq:minima}), and $R_\mathrm{A}^{(1)}>0$ for this well. Respectively, the rate of switching from this well $W_\mathrm{sw} \equiv W_\mathrm{sw}^\mathrm{deeper}$ is smaller than $ W_\mathrm{sw}^{(0)} $ by the factor $\exp(-\alpha_d R_\mathrm{A}^{(1)}/ \lambda)$. The rate of switching from the shallow well is larger than $ W_\mathrm{sw}^{(0)} $ by the inverse of this factor. For concreteness we will consider $R_\mathrm{A}^{(1)}$ for the well at $Q>0$.

\begin{figure}[h]
	\centering
	\includegraphics[width=1\linewidth]{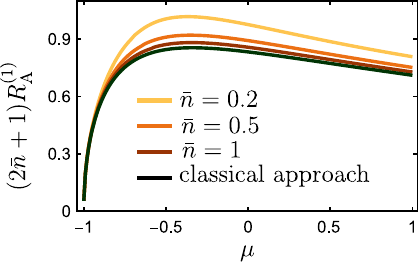}
	\caption{The scaled addition to the quantum activation energy $R_\mathrm{A}^{(1)}$ due to an extra force as given by Eq.~(\ref{eq:def:R_A}). The plot refers to switching from the stable state at $Q>0$ and to the phase of the extra force $\varphi_d = \pi/2$; $\bar n$ is the thermal occupation number of the parametric oscillator, and $\mu = \omega_p(\omega_p-2\omega_0)/F_p$.  }
	\label{fig:log_susc}
\end{figure}

The exponent $R_\mathrm{A}^{(1)}$ for $\varphi_d=\pi/2$ depends on two parameters: the thermal occupation number $\bar n$ and the scaled detuning $\mu$ of  half of the parametric modulation frequency $\omega_p$ from the oscillator eigenfrequency $\omega_0$. The dependence of $R_\mathrm{A}^{(1)}$ on $\mu$ in the range $-1 < \mu < 1$, where the only stable state of the oscillator are period-two vibrational states, is shown in Fig.~\ref{fig:log_susc}. The function $R_\mathrm{A}^{(1)}$ goes to zero near the bifurcation point $\mu=-1$ where the period-two states emerge, see Appendix~\ref{sec:bifurcation}. Overall, the dependence on $\mu$ is nonmonotonic, with a maximum near $\mu= -0.5$. Interestingly, the dependence on $\bar n$ is close to $1/(2\bar n +1)$. In the limit $\bar n\gg 1$ the result approaches the result obtained for the classical regime \cite{Ryvkine2006a}.   

The difference of the switching rates due to the extra force leads to a difference in the stationary populations of the period-two states of the oscillator. For a classical parametrically modulated micromechanical resonator the population difference and its periodic dependence on $\varphi_d$ was observed in Ref.~\cite{Mahboob2010}. 

%%%%%%%%%%%%%%%%%%%%%%%%%%%%%%%%%%%%%%%%%%%%%%%%%%%%%%%%%%%%%%%%%%%%%%

\subsection{High-temperature limit}
\label{subsec:large_bar_n}

It is seen from Eq.~(\ref{eq:transition_rates_Fourier}) that, for a large thermal occupation number $\bar{n}$, the transition probabilities become symmetric,  $\abs{W_{n+m\, n}-W_{n-m\, n}}\ll W_{n+m\, n}$. Then, from Eq.~(\ref{eq:xi_stationary}),  the derivative of the activation energy becomes small, and we can expand $\xi \approx 1 - \omega(g) R'(g) + \left[\omega(g) R'(g)\right]^2/2$. Inserting this expansion into Eq.~(\ref{eq:xi_stationary}) gives 
\begin{align}
\label{eq:R_prime_large_nbar}
R'(g_n) = 2\omega^{-1}(g)\sum_m mW_{n+m\,n}\bigg/\sum_m m^2 W_{n+m\,n}
\end{align}
Using the relation between the transition rates and the Fourier components of $a(\tau,g)$ \cite{Marthaler2006}, the above ratio for the well of $g(Q,P)$ at $Q>0$ can be written as
\begin{align}
\label{eq:R_M_N_general}
&	R'(g) = \frac{2}{2\bar{n}+1} \frac{M(g)}{N(g)}\,, \qquad	M(g) = \iint_{A(g)} d{}{Q} d{}{P}\,,
	\nonumber\\
&	N(g) = \frac{1}{2} \iint_{A(g)} d{}{Q} d{}{P} \,(\partial_Q^2 + \partial_P^2) g(Q,P)\,,
\end{align}
where the integrals run over the interior $A(g)$ of the well limited by the contour $g (Q,P) = g$. The relation (\ref{eq:R_M_N_general}) was derived in Ref.~\cite{Marthaler2006} in the absence of an extra force at half the modulation frequency, but it applies also in the presence of such force.

The correction $\alpha_d R'^{(1)}$ to $R'(g)$ is determined by the corrections $\alpha_d M^{(1)}(g)$ and $\alpha_d N^{(1)}(g)$ to $M(g)$ and  $N(g)$, respectively,
\begin{align}
\label{eq:R_prime_classical}
R'^{(1)}(g) =  \frac{2}{2 \bar{n}+1} \left[\frac{M^{(1)}(g)}{N\0(g)} - \frac{M\0(g)}{N\0{}^2(g)}N^{(1)}(g)\right].
\end{align}
To find these corrections  we note first that $M=2\pi I_f$, where $I_f$ is the full action of the intrawell motion, see Sec.~\ref{subsec:intrawell_dynamics} and Appendix~\ref{subsec:intrawell_frequency}. Then from Eq.~(\ref{eq:first_order_action})
\begin{align}
\label{eq:delta_M}
&M^{(1)} =  \pi \sin \varphi_d
 \end{align}
Using, as in Appendix~\ref{subsec:intrawell_frequency} that, on the trajectory with a given $g=g^{(0)} + \alpha_d g^{(1)}$ and a given $Q$, the correction $\alpha_d P^{(1)}(Q|g)$ to the momentum  is equal to $\alpha_d P^{(1)} = -\alpha_d g^{(1)}/\partial_P g^{(0)}$, one finds
\begin{align}
\label{eq:delta_N}
N^{(1)} = \pi (\mu+2)\sin\varphi_d.
\end{align}
Equations (\ref{eq:delta_M}) and (\ref{eq:delta_N}) give the correction $R'{}\1$ in the explicit form.

An important aspect of the calculation of $R'^{(1)}(g)$ in the limit $\bar n\gg 1$ is that it can be done using in Eq.~(\ref{eq:R_prime_large_nbar}) the general expressions for the  corrections to the transition rates $W^{(1)}_{nn'}$. Comparing the fairly cumbersome expressions for these corrections to the calculation in terms of $M^{(1)}$ and $N^{(1)}$ provides a way to independently check them. The calculation in Appendix~\ref{sec:symmetric} shows that the expressions obtained by two different approaches coincide. 

The change of the activation energy for switching from the well at $Q>0$ is $\alpha_d R_\mathrm{A}^{(1)}$ with
\begin{align}
\label{eq:R_A_classical}
R_A^{(1)} &= \int_{g_\text{min}^{(0)}}^0 R'^{(1)}(g) d{} g + \frac{2 \sin \varphi_d\sqrt{1+\mu}}{(2\bar n +1)(2 + \mu)}
\end{align}
The last term in the above expression is obtained from Eq.~(\ref{eq:R_M_N_general}) by taking into account that the minimum of the $Q>0$-well is shifted from $g_\mathrm{min}\0$ by $-\sqrt{1+\mu}\sin\varphi_d$ and that, for $2\bar n +1  \gg 1$, we have $R'{}\0(g_\mathrm{min}\0) \approx 2[(2+\mu)(2\bar n +1)]^{-1}$.

For large $\bar{n}$ we have $2/\lambda (2\bar{n}+1) \approx F \omega_p^2/6 \gamma k T$. The Planck constant has dropped  out from this expression. The result for $R_\mathrm{A}$ coincides with the expression derived within a classical formulation \cite{Ryvkine2006a}, except that the analytic expression for $M^{(1)}$ and $N^{(1)}$ were not obtained in \cite{Ryvkine2006a}.

%%%%%%%%%%%%%%%%%%%%%%%%%%%%%%%%%%%%%%%

\subsection{Prebifurcation regime}
\label{sec:prebifurcation}

Explicit expressions for $R'$ and for $R_\mathrm{A}$ in the presence of an extra force can be also obtained near the bifurcation point where there emerge the period-two states of the parametrically modulated oscillator. In the absence of dissipation and an extra force, the bifurcation point is $\mu=-1$: for $\mu+1 > 0$ the Hamiltonian function $g\0(Q,P)$ has two minima that correspond to period-two states, whereas for $\mu<-1$ it has one minimum at $Q=P=0$.

Dissipation shifts the position of the bifurcation point, and in a close vicinity to the bifurcation point the oscillator motion is overdamped. The dynamics is controlled by a soft mode, a single dynamical variable that, in the quantum regime, satisfies a first order Langevin equation with the noise intensity $\propto (2\bar n+1)$ \cite{Dykman2007}. We show in Appendix~\ref{sec:bifurcation} that this approach applies also in the presence of an extra force and allows describing the effect of such force on the activation energy of interwell switching.

For weak damping there exists a regime where the oscillator is close, but not too close to the bifurcation point.
In the corresponding parameter range the motion is underdamped, on the one hand  but, on the other hand, the switching rate and the effect of an extra force on this rate display a characteristic scaling  with the distance to the bifurcation point.  We call this a prebifurcation regime, and the corresponding parameter range can be called the prebifurcation range. Where there is no extra force, this range is easy to find by noting that the dimensionless frequency of vibrations about the minimum of $g\0(Q,P)$ is $\omega\0(g\0_\mathrm{min}) = 2(\mu+1)^{1/2}$. The prebifurcation range is where this frequency is small, $\omega\0(g\0_\mathrm{min})\ll 1$, yet it is much larger than the dimensionless decay rate $\kappa$.

For $\omega(g)\ll 1$, one can expand $\exp[-\omega(g)R'(g)]$ in Eq.~(\ref{eq:xi_stationary}), which results in Eq.~(\ref{eq:R_prime_large_nbar}) for $R'(g)$ and ultimately in the expressions (\ref{eq:R_M_N_general}) - (\ref{eq:R_A_classical}) for $R'(g)$ and for the corrections to $R'$ and $R_A$ due to the extra force. We emphasize that these expressions apply even where $\bar n <1$, the only condition is that the system is close to the bifurcation point.  It is easy to show that in the prebifurcation range $N\0(g) \approx M\0(g)$. Taking into account that $g\0_\mathrm{min}=-(\mu +1)^2/4$ we obtain from Eq.~(\ref{eq:R_M_N_general}) $R_\mathrm{A}\0 \approx  (\mu+1)^2/2(2\bar n+1)$ \cite{Marthaler2006}, whereas from Eqs.~(\ref{eq:delta_M}) - (\ref{eq:R_A_classical}) for the well with $Q>0$
\begin{align}
\label{eq:R_A_prebif}
R_\mathrm{A}^{(1)} \approx  \frac{2 \sin \varphi_d }{2\bar{n}+1} \sqrt{\mu+1}, \quad \mu+1 \ll 1.
\end{align}

Interestingly, the correction to the switching rate (\ref{eq:R_A_prebif}) falls off with the decreasing distance to the bifurcation point $\mu +1$ much slower than the leading term $R_\mathrm{A}\0$. This shows that the range of applicability of the perturbation theory shrinks down as the system approaches the bifurcation point. We note that both $R_\mathrm{A}\0$ and $R_\mathrm{A}^{(1)}$ depend on $\bar n$ in the same way, $\propto 1/(2\bar n +1)$. 

It is also interesting that the difference of the values of $R_A^{(1)}$ for the wells with $Q>0$ and $Q<0$ comes only from the difference of their depths. The regions where the RWA energy $g$ in the both wells is the same give equal contributions to $R_A^{(1)}$.

%%%%%%%%%%%%%%%%%%%%%%%%%%%%%%%%%%%%%%%%%%%%%%%%%%%%%%%%%%%%%%%%%%%%%%%%%%%%%%%%%%%%%%%%%%%%%%%

\section{Conclusion}
\label{sec:conclusion}

The results of this paper reveal important aspects of quantum fluctuations in parametric oscillators. The model is well-known and broadly used in quantum and classical physics: a weakly nonlinear oscillator, which is parametrically modulated at frequency $\omega_p$ close to twice the eigenfrequency and additionally driven by a weak force at the frequency $\omega_p/2$.  Without this force, the oscillator dynamics in the frame rotating at frequency $\omega_p/2$ is described by a symmetric double-well Hamiltonian, with the symmetry related to the time shift by the modulation period $2\pi/\omega_p$. The force with twice this period lifts the symmetry. It thus suppresses the tunneling between the symmetric states. One might expect that this would localize the oscillator inside the wells.

The physical picture is qualitatively different in the presence of relaxation. The coupling of the oscillator to a thermal bath leads to dissipation and also to quantum fluctuations. In turn, these fluctuations lead to the inter-well switching in which the oscillator goes over the barrier that separates the wells. This is reminiscent of thermal activation, except that the activation can be caused by quantum fluctuations and can occur for $T=0$.

Our results show that the force at frequency $\omega_p/2$ can {\em exponentially increase} the switching rate. This may be thought of as a reduction of the barrier height. However, the actual process is  more delicate, as the system is far from thermal equilibrium and the conventional picture of quasi-Boltzmann distribution over the intrawell states is inadequate.

Our analysis refers to the case where the wells of the Hamiltonian contain many states, but the decay rate of the oscillator is small, so that the level spacing largely exceeds the level widths. In this case, as we show, the major effect of the extra force is the change of the rates of transitions between the intrawell states. It can be thought of as the change of the random walk over the intrawell states due to quantum fluctuations. Ultimately, this change results in a change of the probability to reach the top of the barrier that separates the wells and then to switch to another well. Because there are many states involved and the effect of the change accumulates, the change of the rate of interwell switching is exponential in the force amplitude. 

In the exponent of the switching rate, the amplitude of the extra force is multiplied by the number of the intrawell states. We find the relevant factor. The general expression simplifies for comparatively large thermal occupation number of the oscillator $\bar n$, in which case the above factor is $\propto (2\bar n +1)^{-1}$. It also simplifies near a bifurcation point, where the factor is shown to scale as the distance to the bifurcation point to the power $1/2$.

The strong effect of the extra force on the rate of switching between the vibrational states of a quantum oscillator suggests a way of an efficient control of such switching. In particular, the possibility to increase the switching rate is important for applications. Our results also provide the means for analyzing the dynamics of networks of coupled quantum parametric oscillators. A major effect of the coupling is the force that vibrations of the coupled oscillators exert on each other. If the oscillators are not identical, such a network presents a quantum nonreciprocal system, since the forces between different oscillators are unbalanced: the force exerted by an oscillator with a larger amplitude on a neighboring oscillator with a smaller amplitude is generally larger than the force exerted back. 

One of the most interesting types of quantum parametric oscillators are vibrational modes in superconducting microcavities with Josephson junctions, which make the cavities nonlinear. Such modes have been intensely studied in the context of cat qubits based on the vibrational states of  parametric oscillators, cf. the recent papers \cite{Reglade2024,Bhandari2024,Venkatraman2024} and references therein. Of importance for suppressing bit-flip errors in the qubits is driving the oscillators to sufficiently large amplitudes, the regime studied in the present paper. An extra force at half the modulation frequency provides a means for controlling the qubits. Thus these qubits are a natural platform for studying the extremely strong and unexpected effects of such force described in this paper.

\begin{acknowledgments}
	D.\,K.\,J.\,B. and W.\,B. gratefully acknowledge financial support from the Deutsche Forschungsgemeinschaft(DFG, German Research Foundation) through Project-ID 425217212 - SFB 1432.	M.I.D. acknowledges partial support  from the US Defense Advanced Research Projects Agency (Grant No. HR0011-23-2-004) and from the Moore Foundation (Grant No. 12214).
\end{acknowledgments}

\appendix 

%%%%%%%%%%%%%%%%%%%%%%%%%%%%%%%%%%%%%%%%%%%%%%%%%%%%%%%%%%%%%%%%%%%%%%

\section{CLASSICAL MOTION}
\label{sec:app_classical_motion}

We calculate the matrix elements $a_m(g_n)= \braket{n+m|\hat a |n}$  in the semiclassical approximation. To this end, we consider the classical motion inside the wells. This is a periodic motion with frequency $\omega(g)$ that depends on the RWA energy $g$. It is described by the Hamiltonian equations
\begin{align}
\label{eq:eom_classical}
	\frac{dQ}{d\tau} &= \frac{\partial g}{\partial P}\,, \qquad \frac{dP}{d\tau} = -\frac{\partial g}{\partial Q}\,.
\end{align}

In the absence of extra force the Hamiltonian of the system is $g^{(0)}$, and we write it as a function of the coordinate $q$ and momentum $p$, i.e., as $g^{(0)}(q,p)$. The Hamiltonian equations for $q,p$ have the form (\ref{eq:eom_classical}) with $g$ replaced with $g^{(0)}$,
\begin{align}
\label{eq:zeroth_order_eom}
dq/d\tau = \partial_pg^{(0)}, \quad dp/d\tau = -\partial_q g^{(0)}.
\end{align}
The solution of these equations for the well at $q>0$  is expressed in terms of the Jacobi elliptic functions \cite{Marthaler2006},
\begin{align}
\label{eq:Jacobi}
	q(\tau,g) &=\frac{2^{3/2} \abs{g}^{1/2} \JacobiDN (\tau' |m_J)}{\varkappa_+ + \varkappa_- \JacobiCN (\tau' | m_J)}\,, \nonumber\\
	p(\tau,g) &=\frac{\varkappa_+ \varkappa_- \abs{g}^{1/4} \JacobiSN (\tau' | m_J)}{\varkappa_+ + \varkappa_- \JacobiCN (\tau' | m_J)}\,,
\end{align}
where
\begin{align*}
	\varkappa_\pm &= (1+\mu \pm 2\abs{g}^{1/2})^{1/2}\,, \qquad \tau' = 2^{3/2} \abs{g}^{1/4} \tau\,, \\
	m_J& \equiv m_J(g,\mu)= \frac{(\mu + 1 -2 \abs{g}^{1/2})(\mu -1 +2 \abs{g}^{1/2})}{8\abs{g}^{1/2}}\,
\end{align*}
(here we use $\varkappa_\pm$ insetad of $\kappa_\pm$ used in \cite{Marthaler2006} to avoid confusion with the relaxation rate parameter $\kappa$).

The Jacobi elliptic functions are double-periodic. The real period is  $\tau_p^{(1)} = 2^{1/2} \abs{g}^{-1/4} K(m_J)$, whereas the second period is complex, $\tau_p^{(2)} = i 2^{1/2} \abs{g}^{-1/4} K(1-m_J)$; here $m_J\equiv m_J(g,\mu)$ is the modulus and $K(m_J)$ is the complete elliptic integral of the first kind. The frequency of the classical motion in the absence of an extra force is $\omega^{(0)}(g) = 2 \pi/\tau_p^{(1)}$. The double-periodicity of the Jacobi elliptic functions allows finding the Fourier components $a_m^{(0)}(g)$ of $a^{(0)}(\tau,g) = \left[p(\tau,g)- i q(\tau,g)\right]/\sqrt{2\lambda}$, i.e., the Fourier components $a_m(g)$ in the absence of the extra force, 
\begin{align}
	\label{eq:fourier_unperturbed}
	&a^{(0)}_m(g) = -i(2\lambda)^{-1/2} \omega\0(g) \frac{\exp(-im\phi_\ast)}{1+\exp(-im\phi_0)}\,, 
	\nonumber\\
	&\phi_0 = \pi \left(1+\tau_p^{(2)} / \tau_p^{(1)} \right)\,,
\end{align}
with $\phi_\ast$ given by the equation
\begin{align*}	
	\JacobiCN(2K\phi_\ast /\pi |m_J) = - \left(\frac{1+\mu +2 \abs{g}^{1/2}}{1+\mu - 2\abs{g}^{1/2}}\right)^{1/2}\,.
\end{align*}
%

%%%%%%%%%%%%%%%%%%%%%%%%%%%%%%%%%%%%%%

\subsection{Frequency of intrawell vibrations}
\label{subsec:intrawell_frequency}

An extra force at $\omega_p/2$ changes the shape of the wells of $g(Q,P)$ and the frequency of the intrawell vibrations. The reciprocal frequency as function of the energy $g$ is given by the derivative of the action $I_f(g)=(2\pi)^{-1}\oint P(Q|g)\,dQ$  over $g$, where $P(Q|g)$ is the momentum on the Hamiltonian trajectory (\ref{eq:eom_classical}) with a given $g$. The action $I_f$ and the momentum $P$ refer to the full time-independent RWA Hamiltonian. Therefore $I_f $ is independent of time, in contrast to the action variable $I(\tau)$ defined for the Hamiltonian $g^{(0)}$.

To the first order in $\alpha_d$, the action $I_f$ is determined by the linear in $\alpha_d$ correction to  the momentum, $P(Q|g)\approx P\0(Q|g) + \alpha_d P^{(1)}(Q|g)$. Since the zeroth-order term $P^{(0)}(Q|g)$ is given by the equation $g^{(0)}(Q,P^{(0)})=g$, from Eq.~(\ref{eq:full_g}) we find  $I_f \approx I_f\0 + \alpha_d I_f^{(1)}$ with
\begin{align}
\label{eq:full_action_correction}
	I^{(1)}_f(g) =&-\frac{1}{2\pi}\oint dQ \,\frac{g^{(1)}\bigl(Q,P^{(0)}(Q|g)\bigr)}{\partial_P g\0},\nonumber\\
	=& \frac{1}{2\pi} \int_{0}^{2\pi/\omega^{(0)}(g)} \left[P^{(0)}(\tau,g) \cos \varphi_d\right. \nonumber\\
&\left.	 + Q^{(0)}(\tau,g) \sin \varphi_d\right]d{}{\tau}\,, 
\end{align}
where $Q\0(\tau,g)=q(\tau,g)$ and $P\0(\tau,g)=p(\tau,g)$ are the dynamical variables in the absence of the extra force described by Eq. (\ref{eq:Jacobi}). Since $p(\tau,g) = - p(-\tau,g)$, the first term in the second line of Eq.~(\ref{eq:full_action_correction}) is zero. The integral over $\tau$ of $q(\tau,g)$ can  be evaluated using the explicit expressions  (\ref{eq:Jacobi}). Alternatively one can write
\[I^{(1)}_f(g) =\frac{1}{\pi} \int q \,dq/\partial_pg^{(0)}(q,p), \]
change  from integration over $q$ to integration over $X = \sqrt{4(g+q^2)+ (\mu-1)^2}$, and use that, with this change, $p^2 + q^2 -(\mu -1) = X$ whereas $q\,dq = X\,dX/4$. Both methods immediately show that, surprisingly, the integral is independent of $g$ and $\mu$, so that the first-order correction to the action  is 
\begin{align}
\label{eq:first_order_action}
	I^{(1)}_f &= \sgn(Q_\mathrm{min})  \sin (\varphi_d)/2\,.
\end{align}
where $Q_\mathrm{min}$ is the position of the minimum of the considered well of $g(Q,P)$. 

From Eq.~(\ref{eq:first_order_action}), $d I^{(1)}_f/dg=0$, and therefore the frequency of intrawell vibrations with a given $g$ is not changed by the extra force, to the first order in the force amplitude. We note that for $g(Q,P) = g_\mathrm{min}$, the overall action is zero. The perturbation theory breaks down for $|\omega^{(0)}(g_\mathrm{min}^{(0)}) (g - g_\mathrm{min}^{(0)})| \sim \alpha_d$.

%%%%%%%%%%%%%%%%%%%%%%%%%%%%%%%%%%%%%%%%%%%%%%%%%%%%%%%%%%%%%%%%%%%%%%%%%%%%%%%%%%%%%%

\section{CORRECTIONS TO THE FOURIER COMPONENTS $a_m$}
\label{sec:corrections_to_a}
In the semiclassical approximation, finding corrections to the transition rates $W_{nn'}$ is reduced to finding corrections to the Fourier components of the functions $a(\tau,g) = (2\lambda)^{-1/2}[P(\tau,g) - iQ(\tau,g)]$. We calculate these corrections perturbatively using the action-angle variables $(I,\psi)$ of the system in the absence of the extra force. This system has the coordinate and momentum $q$ and $p$ and the Hamiltonian function $g^{(0)}(q,p)$. The transformation to $(I,\psi)$ is given by Eq. (\ref{eq:transformation}), which defines for this system $ q(I;\psi), p(I;\psi)$, and $g^{(0)}(I;\psi) \equiv g^{(0)}\bigl(q(I;\psi),p(I;\psi)\bigr) \equiv g^{(0)}(I)$ in terms of $I$ and  $\psi$.  In the presence of the extra force, we defined the coordinate and momentum of the oscillator as functions of $I,\psi$ as $ Q(I;\psi)\equiv q(I;\psi), P(I;\psi)\equiv p(I;\psi)$. The function $a(I;\psi)$ is expressed in terms of the Fourier components $a_m^{(0)}$ as
\begin{align}
\label{eq:a_m_I_components}
a(I;\psi) = \sum_m a_m^{(0)}\bigl(g^{(0)}(I)\bigr)\exp(im\psi).
\end{align}
We remind that we use the notation  $G^{(0)}(I;\psi) \equiv G^{(0)}(I)$ for $g^{(0)}(Q,P)$ expressed in terms of $I,\psi$, cf. Eq. (\ref{eq:G_0}). 

The extra force changes the time evolution of  $I(\tau)$ and $\psi(\tau)$. We find this change from the Hamiltonian equations of motion (\ref{eq:action_angle}) for  the Hamiltonian $G(I;\psi) = G^{(0)}(I) + \alpha_d G^{(1)}(I;\psi)$. The perturbation $G^{(1)}(I;\psi)$ is given by $g^{(1)}\bigl(Q(I;\psi),P(I;\psi)\bigr)$ in Eq.~(\ref{eq:full_g}). Since by construction $Q(I,\psi)$ and $P(I,\psi)$ are periodic in $\psi$, the Hamiltonian $G^{(1)}$ is also periodic in $\psi$. To the leading order in $\alpha_d$,
\begin{align}
	\label{eq:transformation_explicitly}
	G^{(1)}(I;\psi)& = \sum_mG_m^{(1)}(I^{(0)})\exp(im\psi)\,,\nonumber\\ 
	G_m^{(1)}(I) &= - \sqrt{\lambda/2}\left[a_m^{(0)}(g^{(0)}(I)) e^{i \varphi_d} \right. \nonumber\\
	&\left. + \left(a_{-m}^{(0)}(g^{(0)}(I))\right)^* e^{-i\varphi_d} \right]\,. 
\end{align}
Here $I^{(0)}\equiv I^{(0)}(g)$ is given by  the equation  
\[G^{(0)}(I^{(0)})=g;\]
this is the value of $I$ for a given $g$ for $\alpha_d=0$. 

To the first order in $\alpha_d$, the action $I$ has a smooth and oscillating terms. To find the smooth term $\bar I$ for a given energy $g$, following the method of averaging \cite{Arnold1989},  we set $g$ equal to the full period-averaged Hamiltonian $\overline{G(I,\psi)}$,
\begin{align*}
	g = G^{(0)}(\bar{I}) + \alpha_d \bar{G}^{(1)}\,,
\end{align*}
where the bar denotes period averaging, cf. Sec.~\ref{sec:LS_finite_nbar}, so that $\bar{G}^{(1)} = G_0^{(1)}(I^{(0)})$. From this equation we find
\begin{align}
\label{eq:I_bar}
\bar I(g) = I^{(0)}(g) - \frac{\alpha_d}{\omega^{(0)}(g)} G_0^{(1)}(I^{(0)}).
\end{align} 
Here we used $dG^{(0)}/dI = \omega^{(0)}(g)$ for $I=I^{(0)}(g)$.

With the account taken of Eq.~(\ref{eq:I_bar}), the solution of  the equations of motion (\ref{eq:action_angle}) for $I,\psi$ to the first order in $\alpha_d$ for $\overline{G(I,\psi)}=g$ reads
\begin{align}
\label{eq:I_oscillating}
	I = & \bar{I}(g) - \alpha_d \frac{1}{\omega^{(0)}(g)} \sum_{k\neq 0} G_k^{(1)} e^{ik \omega^{(0)}(g) \tau}\,, 
	\end{align}
and
\begin{align}
\label{eq:psi_oscillating}
			\psi = & \omega^{(0)}(g) \tau  
		+ \alpha_d \sum_{k\neq 0} \frac{1}{ik\omega^{(0)}(g)} e^{ik\omega^{(0)}(g) \tau}
		\nonumber\\
		\times &\left[ \frac{d{} G_k^{(1)}}{d{} I}  -\frac{d{}\omega^{(0)}(g)}{d{}g} G_k^{(1)} \right]\,.
	\end{align}
Here $G_k^{(1)}(I)$ and its derivatives are evaluated for $I=I^{(0)}$. 

In deriving the expression for $\psi(\tau)$ we took into account that the term $d{}G^{(0)}/d{}I$ in  Eq.~(\ref{eq:action_angle})  for $d{}\psi/d{}\tau$ has to be calculated for the action given by Eq.~(\ref{eq:I_oscillating}), to the first order in $\alpha_d$. As explained in Sec.~\ref{sec:LS_finite_nbar}, the resulting correction $\propto \alpha_d$  compensates the term $d{}G^{(1)}_0/d{}I$ in $d{} \psi/d{}\tau$, so that  the secular term in $\psi(\tau)$ is $\omega^{(0)}(g)\tau$. Therefore $I(\tau)$ and $\psi(\tau)$ oscillate at frequency $\omega^{(0)}(g)$, to the first order in $\alpha_d$. 

Inserting Eqs.~(\ref{eq:I_oscillating}) and (\ref{eq:psi_oscillating}) into $a(I;\psi)$, we find $a(\tau,g)$ and then the $m\neq 0~$-Fourier components $a_m$ to the first order in $\alpha_d$,
\begin{align*}
&a_m(g) =\frac{\omega^{(0)}(g)}{2\pi}\int_0^{2\pi/\omega^{(0)}} d\tau e^{-im\omega^{(0)}\tau}a(\tau,g)\\
&\approx a_m^{(0)}(g) + \alpha_d a_m^{(1)}(g),
\end{align*}
with
\begin{align}
	\label{eq:a_m_1}
		&a_m^{(1)} (g) = - \sum_k G_k^{(1)} \frac{da_{m-k}^{(0)}}{dg} \nonumber \\
	&+ \sum_{k\neq 0}   a_{m-k}^{(0)} \frac{m-k}{k} \left[\frac{dG_k^{(1)}}{dg} - \frac{d\omega^{(0)}}{dg} \frac{1}{\omega^{(0)}}  G_k^{(1)} \right]\,.
\end{align}
This expression is used in the main text to find the rates of transitions between the intrawell states of the Hamiltonian $g(Q,P)$.

When only the first-order corrections in $\alpha_d$ are taken into account in the transition rates, one should keep in $G_m\1$ only the term $\propto \sin\varphi_d$, as discussed in the main text. Then, since $a_m\0{}^* = -a_m\0$, we have 
\[G_m\1 \to -i\sqrt{\lambda/2}\sin\varphi_d (a_m\0 + a_{-m}\0).\]
This simplifies the numerical calculation of the rates $W_{n+m\,n}\1$ in Eq.~(\ref{eq:rates_corrections}).

%%%%%%%%%%%%%%%%%%%%%%%%%%%%%%%%%%%%%%%%%%%%%%%%%%%%%%%%%%%%%%%%%%%%%%%%%

\section{QUANTUM PERTURBATION THEORY}
\label{sec:q_perturbation_theory}

If the force were weak, with the scaled amplitude ${\alpha_d\ll \lambda}$, corrections to the matrix elements $\braket{n+m|\hat a|n}\approx a_m(g_n)$ could be found by direct perturbation theory. To the first order in $\alpha_d$
\begin{align}
\label{eq:wave_function_1}
&\ket{n} \approx \ket{n}\!^{(0)} + \alpha_d \ket{n}\!^{(1)}, \nonumber\\
&\ket{n}^{(1)}= -\sum_{k \neq 0} \frac{  ^{(0)}\!\!\braket{n+k|g^{(1)}|n}\!^{(0)}}{\lambda k\omega(g_n)}
\ket{n+k}^{(0)}
\end{align}
where $\ket{n}\!^{(0)}$ is the unperturbed wave function of the $n$th intrawell state. Since the intrawell wave functions are nondegenerate, they can be made real functions of $Q$. This is why, since $\hat a = (\hat P-i\hat Q)/\sqrt{2\lambda}$ and $\hat P=-i\lambda\partial_Q$,  the matrix elements $a_m^{(0)} = \braket{n+m|\hat a|n}^{(0)}$ are purely imaginary, cf. Eq.~(\ref{eq:fourier_unperturbed}). 

The term $\propto \cos\varphi_d$ in $\ket{n}^{(1)}$ comes from the term $\propto \hat P=-i\lambda\partial_Q$ in $\hat g^{(1)}$ and therefore is purely imaginary. As a result the corrections $a_m^{(1)}(g_n)$ to the matrix elements of $\hat a$ that come from this term are real and drop out when the real part of the product $a_m^{(0)}{}^*(g_n) a_m^{(1)}(g_n)$ is calculated.

As the analysis of Appendix~\ref{sec:corrections_to_a} shows, this symmetry property holds even where the perturbation is quantum-strong, $\alpha_d\gg \lambda$. 

%%%%%%%%%%%%%%%%%%%%%%%%%%%%%%%%%%%%%%%%%%%%%%%%%%%%%%%%%

\section{LOGARITHMIC SUSCEPTIBILITY FOR CLOSE RATES OF TRANSITIONS UP AND DOWN THE QUASRIENERGY}
\label{sec:symmetric}
In section \ref{sec:LS_finite_nbar} we investigate the high temperature regime and the prebifurcation regime. In these regimes the calculation of the switching rates can be mapped to the classical case \cite{Ryvkine2006a} where the activation energy is given in terms of the integrals $M(g)$ and $N(g)$, see Eq. (\ref{eq:R_M_N_general}). The logarithmic susceptibility is then found in terms of the corrections to these integrals from the extra force. This calculation does not rely on the explicit expression for the matrix elements $a_m$. However, as we show here, one can find the change of the activation energy directly from the corrections to the matrix elements $a_m^{(1)}$ and recover the same result. As indicated in the main text, this provides an important test of the perturbation theory developed in the main text and in Appendix~\ref{sec:corrections_to_a}.

We start by seeking the solution of Eq. (\ref{eq:xi_stationary}) with the same expansion as in the main text
\begin{align*}
	R'(g) &= R'^{(0)}(g) + \alpha_d R'^{(1)}(g)\,.
\end{align*}
With that $R'^{(1)}(g)$ is given by Eq. (\ref{eq:correction_rprime}). In both, the high temperature ($\bar{n} \gg 1 $) and the prebifurcation regime ($ \kappa \ll  (\mu +1 )^{1/2} \ll 1$) the transition rates in the absence of the extra force become almost symmetric, ${|W_{n+m,n}-W_{n-m,n}|\ll W_{n+m,n}}$, that is, the rates of transitions to the states with larger and lower quasienergy are close to each other.

The variable $\xi^{(0)}$, given by the solution of Eq. (\ref{eq:xi_stationary}) in the absence of the extra force, approaches unity as $\omega^{(0)}(g) R'^{(0)}(g)$ goes to zero. We use the expansion
${\xi^{(0)} \approx 1 - \omega^{(0)} R'^{(0)} + (\omega^{(0)} R'^{(0)})^2/2\,,}$
in Eq. (\ref{eq:xi_stationary}) and Eq. (\ref{eq:correction_rprime}) respectively to find
\begin{align}
	R'^{(0)}(g_n) &= \frac{2}{\omega^{(0)}} \sum_m m W_{n+m\, n}^{(0)}/ \sum_m m^2 W_{n+m\, n}^{(0)}\\
	R'^{(1)}(g_n) &= \frac{2}{\omega^{(0)}} \bigg[ \frac{\sum_m m W_{n+m\, n}^{(1)}}{ \sum_m m^2 W_{n+m\, n}^{(0)}} \label{eq:rprime_one_appendix}\\
	& \phantom{= \frac{2}{\omega^{(0)}} \bigg[}- \frac{\sum_m m^2 W_{n+m\, n}^{(1)} \sum_k k W_{n+k\, n}^{(0)}}{\left(\sum_m m^2 W_{n+m\, n}^{(0)} \right)^2}\bigg]\,.\nonumber
\end{align}
The sums involving $W_{n+m\,n}^{(0)}$ have been found in \cite{Marthaler2006} as
\begin{align*}
	\sum_m m W_{n+m,n}^{(0)} &=  \frac{2 \kappa}{2\lambda \pi} M^{(0)}(g_n)\,,\\
	\sum_m m^2 W_{n+m,n}^{(0)} &= 2 \kappa \frac{2\bar{n}+1}{2 \lambda \pi \omega^{(0)}(g)}  N^{(0)}(g_n)\,.
\end{align*}
Here, $M^{(0)}(g)$ and $N^{(0)}(g)$ are expressed in terms of the integrals over the region of $q,p$  limited by the contour $g\0(q,p)=g$ in a given well, see Eq. (\ref{eq:R_M_N_general}). 

The correction $R'^{(1)}$ contains two contributions that are proportional to the two sums, $\mathcal{S}_1$ and $\mathcal{S}_2$:
\begin{align}
\label{eq:sums_for_rates}
	&\sum_m m W_{n+m\,m}^{(1)} = 4 \kappa  \mathcal{S}_1(g_n)\,, \nonumber\\
	&\sum_m m^2 W_{n+m\,m}^{(1)} = 4 \kappa (2 \bar{n} +1)\mathcal{S}_2(g_n)\,, \nonumber\\
	&\mathcal{S}_1(g) =\sum_m m\, \mathrm{Re}\,[a_{-m}^{(0)}{}^*(g) a_{-m}^{(1)}(g) ] \,, \nonumber\\
	&\mathcal{S}_2(g) =\sum_m m^2\, \mathrm{Re}\,[a_{-m}^{(0)}{}^*(g) a_{-m}^{(1)}(g) ] \,. 
\end{align}
We evaluate them by explicitly writing 
\begin{align}
	\label{eq:real_part_correction_appendix}
	\mathrm{Re}\,\big[a_{m}^{(0)}{}^* & a_{m}^{(1)} \big] = - i  \sqrt{\lambda/2}  \sin(\varphi_d) a_m^{(0)} \times \nonumber\\
	\bigg[ \phantom{+}\,&\sum_k \frac{d a_{m-k}^{(0)}}{d g} \left(a_k^{(0)} + a_{-k}^{(0)}\right) \nonumber \\
	+& \sum_{k\neq 0} \frac{d \omega^{(0)}(g)}{dg} \frac{1}{\omega^{(0)}} \frac{m-k}{k} a_{m-k}^{(0)} \left(a_k^{(0)} + a_{-k}^{(0)}\right) \nonumber \\
	-& \sum_{k\neq 0} \frac{m-k}{k} a_{m-k}^{(0)} \left(\frac{d a_k^{(0)} }{d g} + \frac{d a_{-k}^{(0)}}{d g}\right)\bigg]\,.
\end{align}
With that, $\mathcal{S}_{1}$ and $\mathcal{S}_2$ are expressed in terms of three double sums. 

\subsection{Evaluating $\mathcal{S}_1$}
In $\mathcal{S}_1$, the two sums with $k\neq 0$ change signs if the indices of summation are changed as $m\rightarrow m+ k$ and $k \rightarrow -k$. Therefore 
\begin{align}
\label{eq:S_1_simplified_1}
	\mathcal{S}_1 =i \sqrt{\lambda/2} \sin(\varphi_d) \sum_{m,k} m a_m^{(0)} \frac{da_{m-k}^{(0)}}{dg} (a_k^{(0)} + a_{-k}^{(0)})\,.
\end{align}

The matrix elements in this expression are given by the Fourier integrals, 
\begin{align}
	\label{eq:am_integral_appendix}
	a_\ell^{(0)} = \frac{1}{2\pi}\int_0^{2\pi} a^{(0)}(\tau(\phi),g) e^{-i\ell\phi} d \phi\,,
\end{align}
where $a^{(0)}(\tau,g)$ is defined in Appendix \ref{sec:app_classical_motion} and $\phi = \omega^{(0)} \tau$. Substituting this expression into Eq.~(\ref{eq:S_1_simplified_1}) and using that $\sum_m e^{im\phi} = 2\pi \delta(\phi)$ we obtain
\begin{align}
	\label{eq:S_1_simplified}
	\mathcal{S}_1 &= \frac{1}{4\pi \lambda} \sin(\varphi_d) \int_{0}^{2\pi} \frac{q(\tau(\phi),g)}{\omega^{(0)}(g)} d \phi\,,
\end{align}
where we used that $q(\tau,g) = (\lambda/2)^{1/2}[a(\tau,g) + a^*(\tau,g)]$; we remind that $q$ is the coordinate of the oscillator with the Hamiltonian function $g\0(q,p)$. The integral (\ref{eq:S_1_simplified}) can either be evaluated directly as described in Appendix \ref{app_integrals} or by performing the change of variables that is described in Appendix \ref{subsec:intrawell_frequency}. With that,
\begin{align}
	\mathcal S_1 &= \frac{1}{4\pi \lambda} M^{(1)}\,,&
	\sum_m m W_{n+m \, n}^{(1)} &= \frac{2 \kappa}{2\pi \lambda} M^{(1)}\,.
\end{align}
where $M\1$ is given by Eq.~(\ref{eq:M_1_again}).

%%%%%%%%%%%%%%%%%%%%%%%%%%%%%%%%%%%%%%%%%%%

\subsection{Evaluating $\mathcal{S}_2$}
\label{subsec:S_2}

To evaluate $\mathcal{S}_2$ we again use Eq. (\ref{eq:real_part_correction_appendix}). We note that for an arbitrary function $g_k$ that satisfies $g_k = g_{-k}$
\begin{align*}
	\sum_m \sum_{k\neq 0} &m^2 \frac{m-k}{k} a_{m-k}^{(0)} a_m^{(0)} g_{k} \\
	&= \frac{1}{2} \sum_m \sum_{k\neq 0} (m-k)m a_m^{(0)} a_{m-k}^{(0)} g_{k}\,.
\end{align*}
This allows us to rearrange the sums in$\mathcal{S}_2$  where the $k=0$ term is not included by setting $g_k = a_k^{(0)} + a_{-k}^{(0)}$ or the derivative of this expression with respect to $g$, respectively. In this form the $k=0$ contribution is well defined and turns out to be the same for the both sums, except for the sign. This allows us to write 
\begin{align*}
	\mathcal{S}_2	=  -i &\sqrt{\lambda/2} \sin(\varphi_d) \sum_{m,k} \bigg[m^2 a_m^{(0)} \frac{d a_{m-k}^{(0)}}{dg} (a_{k}^{(0)} + a_{-k}^{(0)})\\
	&- \frac{m(m-k)}{2}  a_{m}^{(0)} a_{m-k}^{(0)} \left(\frac{d a_k^{(0)}}{dg} + \frac{d a_{-k}^{(0)}}{g}\right)\\
	&+ \frac{m(m-k)}{2 \omega^{(0)}} \frac{d\omega^{(0)}}{dg}  a_{m}^{(0)} a_{m-k}^{(0)} (a_k^{(0)} + a_{-k}^{(0)}) \bigg]\,.
\end{align*} 

We insert Eq. (\ref{eq:am_integral_appendix}) into these expressions and use partial integration to absorb any prefactors that are proportional to $k$ or $m$. Using again $\sum_m e^{im\phi} = 2\pi \delta(\phi)$, we find the sums and obtain $\mathcal{S}_2$ in the form
\begin{align*}
	\mathcal{S}_2 = &-\frac{\sin \varphi_d}{8 \pi \lambda} \bigg[ \int_{0}^{2\pi}  \bigg(\frac{\partial^2 p}{\partial \phi^2} \frac{\partial p}{\partial g} + \frac{\partial^2 q}{\partial \phi^2} \frac{\partial q}{\partial g}\bigg) 2 q d\phi\\
	+&\int_{0}^{2\pi} \bigg( \bigg(\frac{\partial p}{\partial \phi}\bigg)^2 + \left(\frac{\partial q}{\partial \phi}\right)^2\bigg) \omega^{(0)}(g) \partial_g \bigg(\frac{q(\tau,g)}{\omega^{(0)}(g) }\bigg)	 d\phi\bigg]
\end{align*}
This expression can be further simplified by repeatedly using the relation
\begin{align}
	\frac{\partial p}{\partial \phi} &= -\frac{1}{\omega^{(0)}} \frac{\partial g^{(0)}(q,p)}{\partial q} \,, &  \frac{\partial q}{\partial \phi} &= \frac{1}{\omega^{(0)}} \frac{\partial g^{(0)}(q,p)}{\partial p}\,, \label{eq:derivatives_appendix}
\end{align}
and calculating the second derivatives using explicit forms of the right hand side of Eq.~(\ref{eq:derivatives_appendix}).
This reduces the calculation to the integrals 
\begin{align*}
	\int_0^{2\pi} q(q^2+p^2) d \phi &= \omega^{(0)} \pi (1+\mu)\,,\\
	\int_0^{2\pi} q p^2 d \phi &= \omega^{(0)} \frac{\pi}{8} \left[(1+\mu)^2 + 4g\right]\,,
\end{align*}
which are calculated directly using either the expressions Eq.~(\ref{eq:Jacobi}) or the change of variables described in section \ref{subsec:intrawell_frequency}. This gives 
\begin{align}
	\mathcal S_2 = \frac{1}{4 \pi \lambda \omega^{(0)}} N^{(1)} &= \sin(\varphi_d) \frac{2 + \mu}{4 \lambda \omega^{(0)}} \,,\\
	\sum_m m^2 W_{n+m \, n}^{(1)} &= \frac{2 \kappa (2\bar{n} +1 )}{2\pi \lambda \omega^{(0)}}N^{(1)}(g_n)\,.
\end{align}
With that Eq. (\ref{eq:rprime_one_appendix}) is equivalent to Eq. (\ref{eq:R_prime_classical}).

%%%%%%%%%%%%%%%%%%%%%%%%%%%%%%%%%%%%%%%%%%%%%%%%%%%%%%%%%%%%%%%

\subsection{The integral $M^{(1)}$}
\label{app_integrals}
We can evaluate the appearing integrals explicitly using the solutions of the classical equations. As a simple example we consider  
\begin{align*}
	&\int_{0}^{2\pi} \frac{q(\tau(\phi),g)}{\omega^{(0)}(g)} d \phi\\
	&=2\int_{q_\mathrm{min}}^{q_\mathrm{max}} d{}{q} \frac{q}{\abs{\partial_p g^{(0)}}} = 2 \int_0^{\tau_p^{(1)}/2} q(\tau,g) d{}{\tau}\\
	&= 2(-g)^{1/4} \int_{0}^{2K} \frac{\JacobiDN(t)}{\kappa_+ + \kappa_- \JacobiCN(t)} d{}{t} \\
	&= 2(-g)^{1/4} \int_{1}^{-1} (-1) \frac{1}{\kappa_+ + \kappa_- z} \frac{1}{\sqrt{1-z^2}} d{}{z} \\
	&=\eval{ 2 \arctan(\frac{2(-g)^{1/4}z}{-\kappa_+ - \kappa_- z})}_{1}^{-1} = \pi\,,
\end{align*}
where $t=2^{3/2} (-g)^{1/4} \tau$ and $z = \JacobiCN(t)$. With that
\begin{align}
\label{eq:M_1_again}
	M^{(1)} &=  \pi \sin \varphi_d.
\end{align}
In a similar way also the integrals appearing in appendix \ref{subsec:S_2} can be evaluated. 

%%%%%%%%%%%%%%%%%%%%%%%%%%%%%%%%%%%%%%%%%%%%%%%%%%%%%%%%%%%%%%%%%%%%%%%%%%%%%%%%

\section{NEAR VICINITY OF THE BIFURCATION POINT}
\label{sec:bifurcation}
At $\mu=-1$ the two wells of $g^{(0)}(Q,P)$ merge into a single well with a minimum at $Q=P=0$. Classically, this is a bifurcation point. The vibration frequency at the bottom of the wells of $g\0$ scales as $\omega^{(0)}(g_{min}^{(0)}) = 2 \sqrt{1+\mu}$ for $\mu$ approaching $-1$ from above. It goes to zero as the system approaches the bifurcation point. Eventually the condition of the smallness of the decay rate $\kappa \ll \omega\0(g)$, which is used in the main body of the paper, breaks down, and off-diagonal elements of the density matrix may no longer be disregarded. In this regime we employ a method based on the evolution equation for the Wigner function. 

For a parametrically modulated oscillator in the absence of an extra force, the escape rate close to the bifurcation point was analyzed in \cite{Dykman2007}. The effect on the escape rate of the effective extra force that models the coupling of parametrically modulated oscillators to each other was considered in \cite{Dykman2018}. Here, we consider the effect of a directly applied extra force at half the modulation frequency and use a method different from that used in \cite{Dykman2018}.

Near the bifurcation point the time evolution of the Wigner distribution $\rho_W\equiv \rho_W(Q,P)$ is described by the equation
\begin{align}
\label{eq:rho_W_equation}
	\frac{\partial\rho_W}{\partial\tau} &\approx -\n (\vb{K} \rho_W) + D \nabla^2 \rho_W\,,\\
	\rho_W (Q,P) &= \int d{}{\zeta} e^{i\zeta P/\lambda} \rho\left(Q+ \zeta /2, Q- \zeta/2\right)\,.
\end{align}
Here we use vector notations, the components of the vectors are along the $Q$ and $P$ axes, $\n \equiv (\partial_Q,\partial_P)$. The function $\rho(Q_1, Q_2) = \braket{Q_1|\hat\rho|Q_2}$ is the matrix element of the density matrix on the wave functions in the coordinate representation, $\ket{Q_1} = \delta(Q-Q_1)$.  Equation~(\ref{eq:rho_W_equation}) is obtained for a parametric oscillator in the RWA approximation using the RWA Hamiltonian (\ref{eq:full_g}) and the same model of relaxation as in Eq.~(\ref{eq:balance}). In this model the diffusion constant is 
\[D=\lambda \kappa (\bar{n}+1/2)\] 
(cf.~\cite{Mandel1995}). The drift coefficients are
\begin{align}
\label{eq:drift_Wigner}
	K_Q = \frac{\partial g(Q,P)}{\partial P} - \kappa Q, \quad	K_P = -\frac{\partial g(Q,P)}{\partial Q} - \kappa P.
\end{align}
In deriving Eq.~(\ref{eq:rho_W_equation}) we took into account that, near a bifurcation point, one of the dynamical variables of the system is ``soft''. The distribution over this variable is comparatively broad. This allowed us to  drop the term $\propto \lambda^2$ that contains higher-order derivatives of $\rho_W$ over $Q,P$, cf.~\cite{Dykman2007}.

The term $\propto D$ describes the effect of quantum and classical fluctuations. The bifurcation point is found from the condition that, in the absence of fluctuations, the number of stationary states changes. The positions of the stationary states on the $(Q,P)$ plane are given by the roots of the equations $K_Q=K_P=0$, and it is the number of the real roots of these equations that changes. It is easy to see that, for $\alpha_d=0$,  the value of $\mu$ at the bifurcation point is
\[\mu_B = - \sqrt{1-\kappa^2}.\]
Here we take into account dissipation, which is not small in the near vicinity of the bifurcation point, where $\sqrt{1+\mu}\lesssim \kappa$, and leads to a shift of the bifurcation point from the $\kappa=0$-value. For $\alpha_d=0$ the bifurcation point is located at $Q=P=0$, this is the so-called pitchfork bifurcation.

 For $\alpha_d=0$, to analyze the dynamics near the bifurcation point, where $|\mu -\mu_B|\ll \kappa$, we rotate the coordinate system in the $(Q,P)$-plane from $(Q,P)$ to $(Q',P')$. The rotation angle $\delta$ is given by the relations $\sin2\delta = \kappa$ and $\cos2\delta  = \mu_B$ \cite{Ryvkine2006a}. In the rotated frame, the drift coefficient for the $Q'$-variable, $K_{Q'}$, does not have a linear term $\propto Q'$. This shows that $Q'$ is the ``soft'' mode that slowly varies in time.  The variable $P'$ adiabatically follows this slow variable. One can then follow the general approach \cite{Haken1975} in which one seeks the solution of Eq.~(\ref{eq:rho_W_equation}) on the time scale $\tau \gg 1/\kappa$ in the form of a Gaussian distribution over the ``fast'' variable $P'$ for a given $Q'$. Integration over $P'$ reduces Eq.~(\ref{eq:rho_W_equation}) to an equation for the distribution $\tilde\rho(Q')$ that depends on the single dynamical variable $Q'$.

It is easy to see that the method directly extends to the case where the parametric oscillator is driven by an extra force at frequency $\omega_p/2$, provided $\alpha_d\ll \kappa$. The equation for $\tilde\rho$ in this case reads
\begin{align}
\label{eq:reduced_Wigner}
\frac{\partial\tilde\rho}{\partial\tau} = &\partial_{Q'} \left[\tilde{\rho} \partial_{Q'} U(Q') + D  \partial_{Q'} \tilde{\rho}\right]\,, \nonumber\\
U(Q') = &\frac{\abs{ \mu_B}}{4 \kappa^3} Q'{}^{4} - \frac{\abs{\mu_B}}{2 \kappa} (\mu - \mu_B)  Q'{}^2 
\nonumber\\
& + Q' \alpha_d \cos(\delta + \varphi_d)\,,
\end{align} 

For $\alpha_d=0$ the potential $U(Q')$ depends on the single parameter $\mu - \mu_B$, which is the distance to the bifurcation point along the $\mu$-axis. For $\mu - \mu_B >0$ it is a symmetric double-well potential. The rate of interwell switching is  \cite{Dykman2007}: 
\begin{align}
\label{eq:near_bif_W_0}
	W_\mathrm{sw}^{(0)} &= \frac{\abs{\mu_B} \epsilon}{\sqrt{2}\kappa \pi} \exp(-R_A^{(0)}/\lambda)\,, & R_A^{(0)} &= \frac{\abs{\mu_B} \epsilon^2}{2(2\bar{n}+1)}\,.
\end{align} 
Note that the switching is due to both classical and quantum fluctuations, the switching rate is nonzero for $\bar n=0$. The activation energy $R_A\0$ is just the height of the barrier between the wells of the potential $U(Q')$.

The term $\propto \alpha_d$ tilts the potential. Strictly speaking, for $\alpha_d>0$ instead of a pitchfork bifurcation the oscillator displays a saddle-node bifurcation. We consider the range $\alpha_d\ll |\mu_B|(\mu - \mu_B) ^{3/2}$. Here the oscillator is still far away from the saddle-node bifurcation and has two stable states of parametrically excited vibrations, which correspond to the minima of $U(Q')$. The activation energy of  switching from a given state is again given by the height of the barrier that separates this state from the other state. To the first order in $\alpha_d$, the change of this height $\alpha_d R_A\1$ has opposite signs for the two wells of $U(Q')$. For the well at  $Q>0$ 
\begin{align}
\label{eq:R_A_1_near_bifurcation}
	R_A^{(1)} = - \frac{2 \sqrt{\mu - \mu_B}}{2 \bar{n} + 1} \cos(\delta + \varphi_d)\,.
\end{align}
For vanishing $\kappa$ the angle $\delta \rightarrow \pi/2$ and Eq.~(\ref{eq:R_A_1_near_bifurcation}) coincides with Eq.~(\ref{eq:R_A_prebif}) found in section \ref{sec:prebifurcation}. 

For larger values of $\alpha_d$ the change of the activation energy is no longer linear in $\alpha_d$. As long as the barrier height is large compared to $D$, one can still use the Kramers picture of escape from a potential well. As $\alpha_d$ further increases, this picture becomes inapplicable and effectively one can no longer say that the oscillator has two stable vibrational states: the relaxation time of approaching one of these state becomes indistinguishable from the time of escaping from this state. Importantly, and in distinction from the Kramers analysis, the escape can be due to purely quantum fluctuations.% 

\end{document}